\documentclass[apj]{emulateapj}

\shorttitle{RAM Code}
\shortauthors{Zhang \& MacFadyen}

\begin{document}

\title{RAM: A Relativistic Adaptive Mesh Refinement Hydrodynamics Code}

\author{Weiqun Zhang\altaffilmark{1}}
\affil{Kavli Institute for Particle Astrophysics and Cosmology,
       Stanford University,
       P.O. Box 20450, MS 29, 
       Stanford, CA 94309
}
\email{wqzhang@slac.stanford.edu} 

\and
                                                        
\author{Andrew I. MacFadyen}
\affil{Institute for Advanced Study, Princeton, NJ 08540}
\email{aim@ias.edu}

\altaffiltext{1}{Chandra Fellow}

\begin{abstract}

We have developed a new computer code, RAM, to solve the conservative
equations of special relativistic hydrodynamics (SRHD) using adaptive mesh
refinement (AMR) on parallel computers.  We have implemented a
characteristic-wise, finite difference, weighted essentially non-oscillatory
(WENO) scheme using the full characteristic decomposition of the SRHD
equations to achieve fifth-order accuracy in space.  For time integration we
use the method of lines with a third-order total variation diminishing (TVD)
Runge-Kutta scheme.  We have also implemented fourth and fifth order
Runge-Kutta time integration schemes for comparison.  The implementation
of AMR and parallelization is based on the FLASH code.  RAM is modular and
includes the capability to easily swap hydrodynamics solvers, reconstruction
methods and physics modules.  In addition to WENO we have implemented a
finite volume module with the piecewise parabolic method (PPM) for
reconstruction and the modified Marquina approximate Riemann solver to work
with TVD Runge-Kutta time integration.  We examine the difficulty of
accurately simulating shear flows in numerical relativistic hydrodynamics
codes.  We show that under-resolved simulations of simple test problems with
transverse velocity components produce incorrect results and demonstrate the
ability of RAM to correctly solve these problems.  RAM has been tested in
one, two and three dimensions and in Cartesian, cylindrical and spherical
coordinates.   We have demonstrated fifth-order accuracy for WENO in one
and two dimensions and performed detailed comparison with other schemes for
which we show significantly lower convergence rates.  Extensive testing is
presented demonstrating the ability of RAM to address challenging open
questions in relativistic astrophysics.

\end{abstract}


\section{Introduction}
\label{sec:intro}

Many astrophysical phenomena involve gas moving at relativistic speeds.
Classical sources include active galactic nuclei (AGN), microquasars, pulsar
wind nebulae and gamma-ray bursts (GRBs).  More recently, conclusive evidence
has mounted that a subset of core collapse supernovae are associated with
long duration ($\tau \gtrsim 5$ s) GRBs \citep[e.g.,][]{hjo03,2003dh}.
Astrophysical processes at the endpoint of stellar evolution thus appear
capable of accelerating flows to ultra-relativistic speed ($W \gtrsim 100$,
where $W$ denotes Lorentz factor.)

Additionally, the fading afterglows of cosmological GRBs are now observed
with an increasing rate by the SWIFT satellite \citep{swift}.  GRB afterglows
are produced after the gamma-ray producing relativistic flow transfers its
energy to the circum-burst medium in the form of a strong relativistic shock.
Of particular interest is evidence that the ejecta producing GRBs and their
afterglows are beamed into jets.  The quality of current afterglow
observations require high-resolution multi-dimensional simulations of jetted
relativistic shocks for interpretation.

Relativistic hydrodynamics simulations are more difficult than
Newtonian simulations \citep{nor86}.  However, there has been much
progress in relativistic numerical simulations during the past decade
due to the development of modern numerical methods and the increasing
power of computers.  We refer the reader to a comprehensive review of
numerical relativistic hydrodynamics by \citet{review} and references
therein.  The so-called high-resolution shock-capturing (HRSC)
methods, which are based on the fact that special relativistic
hydrodynamics (SRHD) with causal equation of state is a hyperbolic
system of conservation laws \citep{ani89}, are particularly promising
in modern numerical simulations.  They can achieve very high accuracy
and handle ultra-relativistic flows, strong shocks and contact
discontinuities extremely well.  GENESIS \citep{alo99} is a very
efficient scheme based on HRSC techniques.

The high-order essentially non-oscillatory (ENO) shock capturing schemes
\citep{eno} have been very successful in numerically solving hyperbolic
systems e.g., the Eulerian gas dynamics equations.  ENO-based methods have
previously been implemented for computational relativistic hydrodynamics
\citep{dw95,don98,db02}.  The weighted essentially non-oscillatory (WENO)
schemes \citep[e.g.,][]{liu94,jia96} are recent developments following the
philosophy of ENO schemes.

Adaptive Mesh Refinement (AMR) has played an increasingly important
role in many branches of numerical astrophysics including e.g.,
cosmology, star formation, jets, interacting binaries, stellar wind
collisions, and supernova explosion and remnant evolution
\citep[see][and references therein]{nor04}.  Examples of recent work
in numerical astrophysics include using AMR to solve the equations of
hydrodynamics \citep{pm01}, magnetohydrodynamics (MHD) \citep{bal01},
special relativistic hydrodynamics (SRHD) \citep{hug02}, general
relativistic hydrodynamics \citep{don04} and general relativistic MHD
\citep{cosmos}.

In this paper, we describe a new modular, highly accurate, special
relativistic hydrodynamics code with adaptive mesh refinement, RAM
({\bf R}elativistic {\bf A}daptive {\bf M}esh).  The modular design of
the code allows us to easily change between various algorithms.  We
have implemented the fifth-order WENO scheme of \citet{jia96} in SRHD
for the first time as one of the modules.  We have also implemented an
HRSC module using the modified Marquina flux formula
\citep{marquina,alo99} and piecewise parabolic reconstruction
\citep{ppm} similar to the GENESIS code \citep{alo99}.  We also use
piecewise linear reconstruction (PLM) with both WENO and HRSC and
perform comparisons between all methods for several standard test
cases.  We present detailed tests of the code demonstrating its
ability to handle the extreme resolution requirements of simulating
ultra-relativistic flows.

A primary motivation in writing the RAM code is to study the
relativistic explosions producing cosmological GRBs.  We plan to use
RAM to simulate the relativistic flows produced at the endpoint of
stellar evolution. In particular, numerical simulations of the
collapsar model for GRBs \citep{woosley93,collapsar} are very
challenging because of the ultra-relativistic speed and detailed
micro-physics involved in the problem.  Relativistic hydrodynamical
simulations of jets in collapsars have been done successfully by
several authors \citep{alo00,zwm03,zwh04} using the GENESIS method.
We plan to use RAM to extend these calculations to higher resolution
over a larger dynamic range in lengthscale made possible by AMR.

In \S~\ref{sec:eqns} we present the fundamental equations of SRHD and
describe the conserved variables evolved by our code. In
\S~\ref{sec:schemes} we describe the flux differenced semi-discrete
form of the equations we solve, the algorithms used for construction
of numerical fluxes and how the equations are integrated in time.  In
\S~\ref{sec:test} we present results from standard test problems on a
fixed uniform mesh including convergence tests. In \S~\ref{sec:amr} we
describe the adaptive mesh refinement algorithm employed in RAM by
utilizing components of the FLASH code \citep{flash}, which in turn
adapted the PARAMESH package \citep{paramesh}. In \S~\ref{sec:amrtest}
we present test problems run with AMR in one, two, and three
dimensions including Riemann problems with transverse velocity, a 2D
axisymmetric jet in cylindrical coordinates and a 3D blast wave.  In
\S~\ref{sec:summary} we summarize the results of the paper.  In
Appendix~\ref{app} we provide details of our implementation of SRHD in
curvilinear coordinates.

\section{Governing Equations of Special Relativistic Hydrodynamics}
\label{sec:eqns}

The governing equations of special relativistic hydrodynamics (SRHD)
describe the conservation of rest mass and stress-energy of a fluid:
\begin{equation}
(\rho u^\mu)_{;\mu} = 0,
\end{equation}
and
\begin{equation}
(T^{\mu\nu})_{;\nu} = 0
\end{equation}
where $\rho$ is the rest mass density measured in the fluid frame,
$u^{\mu} = W(c, \mathbf{u})$ is the fluid four-velocity, $W$ is the
Lorentz factor, $c$ is the speed of light, $\mathbf{u}$ is the
classical three-velocity, $T^{\mu\nu}$ is the stress-energy tensor
of the fluid and the subscript $_{;\mu}$ denotes the covariant
derivative.  For a perfect fluid the stress-energy tensor is
\begin{equation}
T^{\mu\nu} = \rho h u^{\mu} u^{\nu} + p g^{\mu\nu},
\end{equation}
where $h \equiv 1 + \epsilon + p/\rho $ is the relativistic specific
enthalpy, $\epsilon$ is the specific internal energy, $p$ is the
pressure and $g^{\mu\nu}$ is the inverse metric which here we take to
be the Minkowski metric though extension to curved spacetimes is
evident.

SRHD can be written as a set of conservation laws \citep[see,
e.g.][]{review},
\begin{equation}
  \frac{\partial{\mathbf{U}}}{\partial{t}} + \sum_{j=1}^{3}
  \frac{\partial{\mathbf{F}^j}}{\partial{x^j}} = 0, \label{dudt}
\end{equation}
where the conserved variable $\mathbf{U}$ is given by
\begin{equation}
  \mathbf{U} = (D, S^1, S^2, S^3, \tau)^{T},
\end{equation} 
and the fluxes are given by
\begin{equation}
  \mathbf{F}^j = (Dv^j, S^{1}v^{j}+p\delta^{j}_{\ 1},
  S^{2}v^{j}+p\delta^{j}_{\ 2}, S^{3}v^{j}+p\delta^{j}_{\ 3}, S^{j}-Dv^j)^{T},
\end{equation}
here $\delta^\mu_{\ \nu}$ is the Kronecker symbol and $v^j$ is the
velocity.  The conserved variables $\mathbf{U}$ include rest mass
density, $D$, the momentum density, $S^j$, and energy density, $\tau$.
They are measured in the laboratory frame, and are given by (assuming
the speed of light $c=1$),
\begin{eqnarray}
     D & = & \rho W \label{d}\\
   S^j & = & \rho h W^2 v^j  \label{sj}\\
  \tau & = & \rho h W^2 - p - \rho W,
\end{eqnarray}
where $j=1,2,3$, $W$ is the Lorentz factor.  The equations~\ref{dudt}
are closed by an equation of state (EOS) given by $p =
p(\rho,\epsilon)$.  For an ideal gas, the EOS reads,
\begin{equation}
  p = (\Gamma - 1) \rho \epsilon,
\end{equation} 
where $\Gamma$ is the adiabatic index of the ideal gas.   

\section{Numerical Schemes}
\label{sec:schemes}

To numerically solve Eq.~\ref{dudt}, each spatial dimension is
discretized into cells.  Using the method of lines, the time dependent
evolution of Eq.~\ref{dudt} can be expressed in the semi-discrete form
\begin{eqnarray}
    \frac{d\mathbf{U}_{i,j,k}}{dt} = 
    - \frac{\mathbf{F}^x_{i+1/2,j,k} -
      \mathbf{F}^x_{i-1/2,j,k}}{\Delta x} \\ \nonumber
    - \frac{\mathbf{F}^y_{i,j+1/2,k} -
      \mathbf{F}^y_{i,j-1/2,k}}{\Delta y} \\ \nonumber
    - \frac{\mathbf{F}^z_{i,j,k+1/2} - \mathbf{F}^z_{i,j,k-1/2}}{\Delta z}, 
    \label{disdudt}
\end{eqnarray}
where $\mathbf{F}^{x}_{i \pm 1/2,j,k}$, $\mathbf{F}^{y}_{i,j\pm
1/2,k}$ and $\mathbf{F}^{z}_{i,j,k\pm 1/2}$ are the fluxes at the
cell interface.

In order to achieve high-order accuracy in time, the time integration
is done using a high-order total variation diminishing (TVD)
Runge-Kutta scheme \citep{shu88}, which combines the first-order
forward Euler steps and involves prediction and correction.  For
example, the third-order accuracy can be achieved via
\begin{eqnarray}
  \mathbf{U}^{(1)} & = & \mathbf{U}^n + \Delta{t} L(\mathbf{U}^n) 
        \label{eq:rk3:1} \\
  \mathbf{U}^{(2)} & = & \frac{3}{4}\mathbf{U}^n +  
       \frac{1}{4}\mathbf{U}^{(1)} + \frac{1}{4} \Delta{t}
       L(\mathbf{U}^{(1)}) \label{eq:rk3:2} \\
  \mathbf{U}^{n+1} & = & \frac{1}{3}\mathbf{U}^n +
       \frac{2}{3}\mathbf{U}^{(2)} + \frac{2}{3}\Delta{t}
       L(\mathbf{U}^{(2)}) \label{eq:rk3:3},
\end{eqnarray}
where $L(\mathbf{U})$ is the right hand side of Eq.~\ref{disdudt},
$\mathbf{U}^{n+1}$ is the final value after advancing one time step
from $\mathbf{U}^{n}$.  Besides the third-order Runge-Kutta method
(RK3), the standard fourth-order Runge-Kutta (RK4) and fifth-order
Runge-Kutta method (RK5) \citep{rk5} can also be used for time
integration.  We usually use TVD RK3 in our calculations unless
otherwise stated.  We have also implemented and tested RK4 and RK5 and
use them in some calculations.

Note the method of lines, treats information from corner zones
differently than methods using time-averaged fluxes which make some
use of corner information when implemented with Strang splitting.
Recently, dimensionally unsplit methods have been developed
\citep{mig05,gs05} which use the corner-transport upwind (CTU) method of
\citep{ctu} to include corner information completely.

RAM, however, uses the third-order Runge-Kutta algorithm for time integration
consisting of several forward Euler substeps which are corrected to achieve
high accuracy.  Since the final results after one full time step are computed
through several substeps, the corner cells (e.g.,
($x_{i+1},y_{j+1},z_{k+1}$)) of a zone at ($x_{i},y_{j},z_{k}$) affect its
evolution in a full step, although this is not the case for an individual
substep (\S~\ref{sec:f-x} \& \S~\ref{sec:u-x}).  In this sense, RAM, is
dimensionally unsplit.  Comparison with codes using CTU methods on
multi-dimensional test problems would be of value to perform in the future.
We note that multi-dimensional test problems we have run with RAM preserve
symmetries present in the initial conditions (see, e.g.,
Fig.~\ref{fig:riemann2d}).

To solve the fluxes across the cell interfaces, we have implemented
two schemes.  Both schemes involve a reconstruction step to compute
the variables at the cell interfaces.  In the first class of schemes
(\S~\ref{sec:f-x}), the reconstruction is carried out on fluxes.  The
interface flux is obtained by,
\begin{equation}
 \mathbf{F}_{i+1/2} =
  F(\mathbf{F}_{i-r},...,\mathbf{F}_{i+s}),
\end{equation}
where the stencil $(i-r,i-r+1,...,i+s)$ depends on the choice of the
reconstruction scheme.  This class can be considered finite
difference schemes.  We use F-X to denote this class of schemes, where
X stands for the reconstruction scheme.  In the other class of schemes
(\S~\ref{sec:u-x}), the reconstruction is carried out on the unknown
variables.  Then the interface unknowns, $\mathbf{U}_{i+1/2}^{-,+} =
U(\mathbf{U}_{i-r},...,\mathbf{U}_{i+s})$, are used to obtain
interface fluxes,
\begin{equation}
 \mathbf{F}_{i+1/2} = R(\mathbf{U}_{i+1/2}^{-}, \mathbf{U}_{i+1/2}^{+}),
\end{equation} 
where $-$ and $+$ denote the left and right side of the interface
$i+1/2$, respectively.  The flux function $R$ is an approximate
Riemann solver.  The cell unknowns $\mathbf{U_i}$ are considered as
cell averages.  Thus, the second class can be considered as finite
volume schemes.  We use U-X to denote the second class of schemes,
where X stands for the reconstruction scheme.

\subsection{F-X Schemes: Reconstruction Of Fluxes}
\label{sec:f-x}

In the F-X schemes, we have implemented the weighted essentially
non-oscillatory (WENO) scheme and piecewise linear method (PLM) for
the reconstruction of fluxes.  

Essentially non-oscillatory (ENO) finite difference schemes
\citep{eno} for equations of hyperbolic conservation laws use adaptive
stencils in calculating the fluxes across the cell interfaces so that
high-order accuracy can be achieved and numerical oscillations near
discontinuities can be significantly reduced.  Later total variation
diminishing (TVD) Runge-Kutta methods were applied to ENO schemes to
make the schemes more computationally efficient, and the ENO
reconstructions were carried out on fluxes instead of cell averages
\citep{shu88, shu89}.  ENO schemes have been further improved by many
researchers \citep[see][for a review]{shu97}.  One improved scheme is
the fourth-order weighted essentially non-oscillatory (WENO) scheme
which was first introduced by \citet{liu94}.  In the WENO schemes, a
weighted combination of several possible stencils are used instead of
just one stencil.  This would improve the accuracy while keeping the
essentially non-oscillatory property near discontinuities.  A new
algorithm for computing the weights has been used in a modified
fifth-order WENO scheme developed by \citet{jia96}.  The WENO scheme
of \citet{jia96} has been applied by \citet{fen04} to cosmological
simulations.  In this paper, we have implemented the fifth-order
modified WENO scheme of \citet{jia96} in special relativistic
hydrodynamics.  The WENO schemes for hyperbolic conservation laws have
various forms \citep{shu97}.  The particular WENO scheme we use is the
characteristic-wise flux splitting finite difference scheme.

In the WENO scheme, the partial differential equations are discretized
into cells in spatial dimensions (Eq.~\ref{disdudt}).  The evolution
of the partial differential equations is solved by using a TVD
Runge-Kutta scheme.  The key component of the WENO scheme is to
compute the fluxes at the cell interfaces.  As an example, we shall
consider the $x$-direction flux across the cell interface between
$x_i$ and $x_{i+1}$.  We shall assume that states at $x_{i-2}$,
$x_{i-1}$, $x_{i+2}$, and $x_{i+3}$ are either in the computational
domain or can be supplied by boundary conditions.

First, we construct a Roe type average state at the cell interface
$x_{i+1/2}$ from the left state at $x_i$ and right state at $x_{i+1}$.
The eigenvectors of the average state will be used as the basis for
the decomposition of fluxes, $\mathbf{F}_m$, where
$m=i-2,i-1,i,i+1,i+2,i+3$.  We use $\sqrt{\rho h}$ as weight
\citep{eul95} to compute the weighted averaged of pressure, density,
and velocity.  Strictly speaking, this does not satisfy all of the
Roe's conditions \citep{roe81}.  However, we do not use the average
state to compute the flux directly.  Therefore strict Roe's conditions
are not required.  Indeed, simply taking arithmetic averages of the
primitive variables also works very well.

For special relativistic hydrodynamics equations, the complete
characteristic structure of these hyperbolic equations have been
derived by \citet{don98}.  Thus it is feasible to extend the
characteristic-wise WENO to special relativistic hydrodynamics knowing
the eigenvalues and eigenvectors of the Jacobian matrices of the
equations.  The eigenvalues of the Jacobian matrices,
\begin{equation}
\mathbf{B} =
\frac{\partial{\mathbf{F}(\mathbf{U})}}{\partial{\mathbf{U}}}, 
\end{equation}
namely the speed of characteristic waves, can have different signs.
In other words, these waves could propagate towards different
directions.  The second step of the WENO scheme is to split the flux
into left-going and right-going fluxes so that we can treat them
separately.  This flux splitting approach can avoid entropy violation
and make the scheme more robust.  We use the local Lax-Freiderichs
splitting given by, 
\begin{eqnarray}
\mathbf{F}^{+} = \mathbf{F} + \alpha \mathbf{U}\ \\
\mathbf{F}^{-} = \mathbf{F} - \alpha \mathbf{U},
\end{eqnarray}
where $\alpha$ is the local maximum of the absolute values of wave
speed for states at $x_{i-2,i-1,i,i+1,i+2,i+3}$.  The eigenvalues are
all positive for the right-going flux $\mathbf{F}^{+}$, whereas they
are negative for the left-going flux $\mathbf{F}^{-}$.

Because of the obvious symmetry between the left-going flux
$\mathbf{F}^{-}$ and right-going flux $\mathbf{F}^{+}$, we shall only
consider here the right-going waves.  The stencil for the right-going
waves at $i+1/2$, which consists of $x_{i-2,i-1,i,i+1,i+2}$ in the
fifth-order WENO scheme, is slightly upwind-biased.  We can expand
$\mathbf{F}^+_{m}$, where $m=i-2,i-1,i,i+1,i+2$, in terms of the five
right eigenvectors \citep{don98} of the average state at $i+1/2$,
$\mathbf{R}_{n}$, where $n=1,2,3,4,5$.  The expansion is given by,
\begin{equation}
  \mathbf{F}^+_{m} = \sum_{n=1}^{5}{c_{mn} \mathbf{R}_{n}},
\end{equation}
where, $m =i-2,i-1,i,i+1,i+2$.  It is very straightforward to compute
$c_{mn}$ since we also know the left eigenvectors, $\mathbf{L}_n$, of
the Jacobian matrix.  $c_{mn}$ is given by,
\begin{equation}
  c_{mn} = \mathbf{L_n} \cdot \mathbf{F}^+_{m}.
\end{equation}

Given $c_{mn}$, we can construct the coefficients $c_{i+1/2,n}$ for
the cell interface at $x_{i+1/2}$ by giving different weights to
different $c_{mn}$.  The weights depend upon the smoothness of the
stencils, and smooth stencils are given more weight.  This results in
the fifth-order accuracy in smooth region and ENO near
discontinuities.  For the details of how the weights are chosen in the
WENO reconstruction scheme, we refer the readers to the work of
\citet{jia96}.

Using the coefficients $c_{i+1/2,n}$, where $n$ denotes characteristic
waves, the right-going flux at the cell interface at $x_{i+1/2}$ is
given by,
\begin{equation}
   \mathbf{F}^{+}_{i+1/2} = \sum^{5}_{n=1}{c_{i+1/2,n} \mathbf{R}_{n}}.
\end{equation}
Because of the obvious symmetry, the left-going flux,
$\mathbf{F}^{-}_{i+1/2}$, can be computed in the similar procedure.
Note that the left-going flux is also upwind-biased, namely
right-biased. 

In computing the WENO weights, a parameter, $\epsilon$, is introduced
to avoid denominator being zero.  This is the only free parameter in
the WENO scheme.  Moreover, the results are insensitive to the value
of $\epsilon$ as long as it is a small number about $10^{-6}$.

We use F-WENO to denote the above method, which uses the WENO scheme
for the characteristic-wise reconstruction of splitted fluxes.  The
reconstruction of the coefficients $c_{i+1/2,n}$ from $c_{mn}$ can
also be computed using other high-order reconstruction algorithms
instead of the WENO reconstruction algorithm of \citet{jia96}.
Besides the WENO, we have also implemented the PLM for the
reconstruction with a generalized minmod slope limiter \citep{kur00}.
Given $c_{i-1}$, $c_i$, and $c_{i+1}$, the left-biased interface value
reads,
\begin{eqnarray}
  c_{i+1/2} = c_i + 0.5 \mathrm{minmod}( \theta (c_i-c_{i-1}), \\ \nonumber 
  0.5 (c_{i+1}-c_{i-1}), \theta (c_{i+1}-c_i) ),
\end{eqnarray}
where $1 \le \theta \le 2$, and the minmod function reads,
\begin{eqnarray}
\mathrm{minmod}(x,y,z) = \frac{1}{4} (\mathrm{sgn}(x) +
\mathrm{sgn}(y)) \\ \nonumber
   (\mathrm{sgn}(x) + \mathrm{sgn}(z)) \min(|x|,|y|,|z|),
\end{eqnarray}  
here the $\mathrm{sgn}$ function returns the sign of the number.  This
becomes the more diffusive normal minmod limiter when $\theta = 1$,
and becomes the monotonized central-difference limiter of
\citet{mclimiter} when $\theta = 2$.  We usually use $\theta = 1.5$
unless otherwise stated.  We call this method F-PLM.  Our numerical
tests show that the results of F-PLM are comparable to those of F-WENO
(\S~\ref{sec:test}).

\subsection{U-X Schemes: Reconstruction Of Unknowns}
\label{sec:u-x}

Besides the F-X schemes (\S~\ref{sec:f-x}), we have also implemented
another method of computing the flux at the interface to work with the
Runge-Kutta scheme for time integration of Eq.~\ref{disdudt}.  In this
class of schemes, U-X, instead of reconstructing flux directly as in
the F-X schemes, a left state and a right state at the interface are
reconstructed, and then the flux at the interface is calculated using
these two states.  A lot of recent methods, including the so-called
high-resolution shock-capturing (HRSC) methods like the GENESIS method
\citep{alo99} and some high-resolution central schemes
\citep{ls04,db02}, can be considered to be in this category.  Their
main difference is how to compute the flux at the interface given the
left and right states and how to reconstruct interface states.  In the
HRSC methods, an approximate Riemann solver utilizing the information
of the characteristic waves is used, whereas the central schemes use
simplified expressions for the flux without using the characteristic
information beyond the fastest wave speed, which is nevertheless
required for the Courant-Friedrich-Levy (CFL) condition.

The reconstruction of states at the cell interfaces is usually performed by a
high-order interpolation method like the piecewise parabolic method (PPM)
\citep{ppm} or piecewise linear method (PLM).  We have implemented both
reconstruction methods in the RAM code.  For the parameters in the PPM
algorithm, we choose the values suggested by \citet{rppm} for almost all the
tests with PPM.  We consider it undesirable to fine-tune these parameters
to achieve better agreement with analytic solutions for specific tests.
Doing so, we believe, can create a false sense of how a given code performs
in general simulations for which the exact solution is not previously
known.   For the PLM algorithm, we use a generalized minmod slope limiter
as described in \S~\ref{sec:f-x}.  Again, the parameter $\theta = 1.5$ is
used by default.  The reconstruction is performed on the so-called primitive
variables instead of conserved variables because unphysical conserved
variables may arise otherwise.  The pressure and the proper density are
reconstructed directly, whereas velocities are reconstructed using a
combination of reconstructing three-velocity and Lorentz factor.  Since both
the velocities and Lorentz factor are reconstructed, they are unlikely to be
self consistent.  To make them consistent, four-velocities are derived by
multiplying three-velocities with Lorentz factor.  Using these
four-velocities, new consistent velocities and Lorentz factor could be
derived.  We found that this procedure is usually more robust than using only
three-velocities or only four-velocities for the construction of velocity and
Lorentz factor.

To compute the flux across the cell interface given the reconstructed
right and left states at the interface, we have implemented several
Riemann solvers including the modified Marquina's flux \citep{alo99},
local Lax-Friedrichs flux and relativistic HLLE \citep{sch93} in the
RAM code.  A comparison of several schemes for computing the flux has
been performed by \citet{ls04}.  They have shown that the results are
relatively independent of the flux schemes.  For the results shown in
this paper, the modified Marquina's algorithm is used to compute the
interface fluxes.

\subsection{Failsafe Time Integration and Root Finder}
\label{sec:fallback}

Since our code is explicit, the time step for integration is subject
to the Courant-Friedrich-Levy (CFL) condition.  We usually choose a
CFL number to be $0.2$-$0.5$.  Unfortunately, it is not always
failsafe.  Occasionally unphysical results can be produced for
ultra-relativistic flows unless a very small CFL number or a diffusive
scheme is used.  But a very small CFL number is sometimes unacceptable
because of the time it costs.  To make the code robust, a fall-back
mechanism is employed in our code, though we found that such failures
rarely occur.  That is a normal CFL number is used for the first try
of time integrations.  If it fails, the code will return to the
beginning of the failed step and use more diffusive schemes for
reconstruction and/or use a smaller time step.  The recalculation with
more diffusive schemes is performed on the whole grid, when AMR is not
being used.  The fall-back method on AMR grids will be further
discussed in \S~\ref{sec:amr}.  The fall-back requires us to keep a
copy of the original states of conserved variables for a period of the
whole Runge-Kutta step in order to be able to fall back.  Fortunately,
the Runge-Kutta scheme we use \citep{shu88} requires the original
states to be saved anyway
(Eqns.~\ref{eq:rk3:1},~\ref{eq:rk3:2},~\&~\ref{eq:rk3:3}).  Thus it
does not increase the memory usage of the computation.  In the
subsequent steps, the normal reconstruction scheme will be used and
the CFL number will gradually increase to its normal value if it was
previously decreased.  The fall-back mechanism makes the code almost
failsafe. If unphysical results persist even when the first-order
method is employed and a very small time step is used, the numerical
simulation will stop.  One possible remedy when the code runs out of
methods is to set the pressure of bad cells to a floor value.
Fortunately this has never happened in our simulations.

In each time step, conserved variables $\mathbf{U} = (D, S^1, S^2,
S^3, \tau)^{T}$ are updated directly.  In order to calculate the
fluxes at the cell interfaces, physical variables: pressure, proper
rest mass density and velocity need to be computed.  The processes of
recovering physical variables from conserved variables involve a
quartic equation for velocity \citep[e.g.,][]{dun94}.  Though the
quartic can be solved analytically, computing the analytic solution is
actually more expensive than a simple numerical root finder, such as
Newton-Raphson iteration.  Before the Newton-Raphson iteration,
a certain condition, $(\tau + D)^2 > D^2 + S^2$, is checked to make
sure that physical results can be produced from given conserved
variables.  Sometimes, unphysical results, such as negative pressure
and large velocity $v>1$, will be produced.  Then the fall-back
mechanism will be used until physical results are produced or the
maximum allowed number of fall-back is reached.

\section{Test Problems}
\label{sec:test}

For any numerical code, it is very important to do substantial tests.
We have done a series of tests with our code.  Some of the tests are
the so-called Riemann problems, which consist of computing the decay
of an initial discontinuity of two constant states.  It is possible to
get exact solutions for relativistic Riemann problems \citep{exact}.
We have also done some extensively studied problems which have no
analytic solutions.  In all cases, our results are comparable to
published results.  In this section, we will present our numerical
results for some standard tests.  We will compare four schemes
(\S~\ref{sec:schemes}): the finite difference characteristic-wise
WENO, the finite difference characteristic-wise PLM, the finite volume
component-wise PPM and the finite volume component-wise PLM, denoted
by F-WENO, F-PLM, U-PPM, and U-PLM, respectively.  A CFL number of 0.5
is used for these tests unless otherwise stated.

\subsection{One-Dimensional Riemann Problem 1}

\begin{figure}
%\epsscale{1.0}
%\plotone{rie1d1.eps}
\plotone{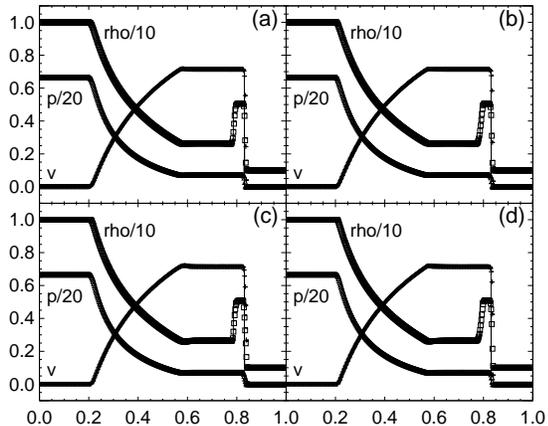}
\caption{ One-dimensional Riemann problem 1 at $t=0.4$.  Results for
  four schemes: (a) F-WENO; (b) F-PLM; (c) U-PPM; (d) U-PLM are shown.
  The computational grid consists of 400 zones.  Numerical results are
  shown in symbols, whereas the exact solution is shown in solid
  lines.  We show proper mass density (square), pressure (triangle)
  and velocity (plus sign).
\label{fig:rie1d1}}
\end{figure}

In this test, the one-dimensional numerical region ($0 \le x \le 1$)
initially consists of two constant states: $p_L = 13.33$, $\rho_L =
10.0$, $v_L = 0.0$ and $p_R = 10^{-8}$, $\rho_R = 1.0$, $v_R = 0.0$,
where $L$ stands for the left state, and $R$ the right state.  The
fluid is assumed to be an ideal gas with an adiabatic index $\Gamma =
5/3$.  The initial discontinuity is at $x = 0.5$.  The results are
shown in Fig.~\ref{fig:rie1d1}.  In this test problem, the evolution
of the initial discontinuity gives rise to a shock, a rarefaction
wave, and a contact discontinuity.  This is a fairly easy test.  All
modern special relativistic hydrodynamics codes can capture the
expected features, acquire correct positions of the shock front,
contact discontinuity and rarefaction wave.  To quantitatively measure
the errors, we calculated the $L_1$ norm errors, $L_1 = \sum_j{\Delta
  x_j|u_j-u(x_j)|}$, where $u(x_j)$ is the exact solution at $x_j$ and
$u_j$ is the numerical result.  The $L_1$ norm errors of density for
four schemes with various grid resolutions at $t=0.4$ are shown in
Table~\ref{tab:rie1d1}.  The accuracy of our results is comparable to
that of \citet{ls04,rppm}.  The order of the convergence rate is about
$1$.  This is consistent with the fact that there are discontinuities
in the solution. 

% f1

% tab 1
\begin{deluxetable}{cccc}
%\tabletypesize{\scriptsize} 
\tablecaption{$L_1$ errors of the density for the 1D Riemann Problem
  1.  Four schemes with various resolutions with uniform spacing are
  shown at $t = 0.4$.
\label{tab:rie1d1}}
\tablewidth{0pt}
\tablehead{
\colhead{Scheme}    &   \colhead{$N$\tablenotemark{a}}  &  
\colhead{$L_1$ Error} & \colhead{Convergence Rate}
}
\startdata
F-WENO & 100  & 1.31e-1 &      \\ 
       & 200  & 7.25e-2 & 0.85 \\ 
       & 400  & 3.32e-2 & 1.1  \\ 
       & 800  & 2.08e-2 & 0.67 \\
       & 1600 & 1.00e-2 & 1.1  \\ 
       & 3200 & 5.07e-3 & 0.98 \\ 
\tableline
F-PLM  & 100  & 1.47e-1 &      \\ 
       & 200  & 8.50e-2 & 0.79  \\ 
       & 400  & 4.06e-2 & 1.1  \\ 
       & 800  & 2.33e-2 & 0.80 \\ 
       & 1600 & 1.22e-2 & 0.93 \\ 
       & 3200 & 7.48e-3 & 0.71 \\ 
\tableline
U-PPM  & 100  & 1.27e-1 &      \\ 
       & 200  & 7.30e-2 & 0.80 \\ 
       & 400  & 3.47e-2 & 1.1  \\ 
       & 800  & 1.97e-2 & 0.82 \\ 
       & 1600 & 9.77e-3 & 1.0  \\ 
       & 3200 & 5.10e-3 & 0.94 \\ 
\tableline
U-PLM  & 100  & 1.32e-1 &      \\ 
       & 200  & 8.57e-2 & 0.62 \\ 
       & 400  & 3.86e-2 & 1.2  \\ 
       & 800  & 2.27e-2 & 0.77 \\ 
       & 1600 & 1.15e-2 & 0.98 \\ 
       & 3200 & 6.48e-3 & 0.83 
\enddata
\tablenotetext{a}{Number of grid points}
\end{deluxetable}

\subsection{One-Dimensional Riemann Problem 2}
\label{sec:rie1d2}

\begin{figure}
%\epsscale{1.0}
%\plotone{rie1d2.eps}
\plotone{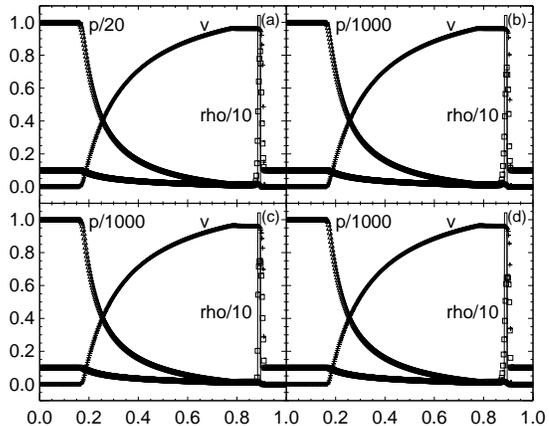}
\caption{ One-dimensional Riemann problem 2 at $t=0.4$.  Results for
  four schemes: (a) F-WENO; (b) F-PLM; (c) U-PPM; (d) U-PLM are shown.
  The computational grid consists of 400 zones.  Numerical results are
  shown in symbols, whereas the exact solution is shown in solid
  lines.  We show proper mass density (square), pressure (triangle)
  and velocity (plus sign).
\label{fig:rie1d2}}
\end{figure}

In this test, the one-dimensional numerical region ($0 \le x \le 1$)
initially consists of two constant states: $p_L = 1000.0$, $\rho_L =
1.0$, $v_L = 0.0$ and $p_R = 10^{-2}$, $\rho_R = 1.0$, $v_R = 0.0$,
where $L$ stands for the left state, and $R$ the right state.  The
fluid is assumed to be an ideal gas with an adiabatic index $\Gamma =
5/3$.  The initial discontinuity is at $x = 0.5$.  The results are
shown in Fig.~\ref{fig:rie1d2}.  In this test problem, the evolution
of the initial discontinuity gives rise to a right-moving shock, a
left-moving rarefaction wave, and a contact discontinuity in between.
Behind the shock, there is an extremely thin dense shell.  The width
of the shell is only $0.01056$ at $t=0.4$.  With 400 uniform zones for
$0 \le x \le 1$, the thin shell is only covered by $4.2$ zones.  Due
to the smearing at the contact discontinuity and the shock, it is not
surprising that the thin shell is not well resolved in our results
with only 400 zones.  At this resolution the maximum density in the
shell for F-WENO, F-PLM, U-PPM and U-PLM is 79\%, 69\%, 72\% and 63\%
of the analytic value.  However, there is no difficulty in resolving
the thin shell with increased resolution.  We have calculated the
$L_1$ norm errors of density for various numerical resolutions (see
Table \ref{tab:rie1d2}) and achieve convergence as expected for
problems with sharp discontinuities.  The results are consistent with
other published results \citep[e.g.,][]{ls04}. 

% f2

% tab 2
\begin{deluxetable}{cccc}
%\tabletypesize{\scriptsize} 
\tablecaption{$L_1$ errors of the density for the 1D Riemann Problem
  2.  Four schemes with various resolutions with uniform spacing are
  shown at $t = 0.4$.
\label{tab:rie1d2}}
\tablewidth{0pt}
\tablehead{
\colhead{Scheme}    &   \colhead{$N$\tablenotemark{a}}  &  
\colhead{$L_1$ Error} & \colhead{Convergence Rate}
}
\startdata
F-WENO & 100  & 2.10e-1 &      \\ 
       & 200  & 1.42e-1 & 0.56 \\ 
       & 400  & 9.29e-2 & 0.61 \\ 
       & 800  & 5.54e-2 & 0.75 \\
       & 1600 & 2.54e-2 & 1.1  \\ 
       & 3200 & 1.51e-2 & 0.75 \\ 
\tableline
F-PLM  & 100  & 1.96e-1 &      \\ 
       & 200  & 1.42e-1 & 0.46  \\ 
       & 400  & 1.06e-1 & 0.42 \\ 
       & 800  & 7.21e-2 & 0.56 \\ 
       & 1600 & 3.92e-2 & 0.88  \\ 
       & 3200 & 2.44e-2 & 0.68 \\ 
\tableline
U-PPM  & 100  & 2.18e-1 &      \\ 
       & 200  & 1.52e-1 & 0.52 \\ 
       & 400  & 9.52e-2 & 0.68 \\ 
       & 800  & 5.42e-2 & 0.81 \\ 
       & 1600 & 2.67e-2 & 1.0  \\ 
       & 3200 & 1.67e-2 & 0.68 \\ 
\tableline
U-PLM  & 100  & 2.13e-1 &      \\ 
       & 200  & 1.65e-1 & 0.37 \\ 
       & 400  & 1.25e-1 & 0.41 \\ 
       & 800  & 8.68e-2 & 0.53 \\ 
       & 1600 & 4.49e-2 & 0.95 \\ 
       & 3200 & 2.71e-2 & 0.73 
\enddata
\tablenotetext{a}{Number of grid points}
\end{deluxetable}

\subsection{One-Dimensional Riemann Problem 3}

\begin{figure}
%\epsscale{1.0}
%\plotone{rie1d3.eps}
\plotone{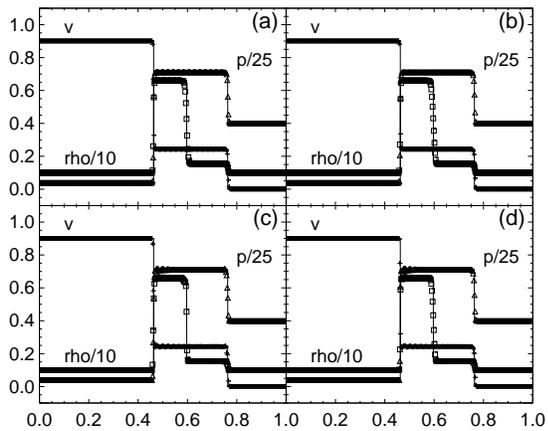}
\caption{ One-dimensional Riemann problem 3 at $t=0.4$.  Results for
  four schemes: (a) F-WENO; (b) F-PLM; (c) U-PPM; (d) U-PLM are shown.
  The computational grid consists of 400 zones.  Numerical results are
  shown in symbols, whereas the exact solution is shown in solid
  lines.  We show proper mass density (square), pressure (triangle)
  and velocity (plus sign).  This problem contains a strong reverse
  shock in which post-shock pressure oscillations are visible in
  panels (c) and (d).
\label{fig:rie1d3}}
\end{figure}

In this test, the one-dimensional numerical region ($0 \le x \le 1$)
initially consists of two constant states: $p_L = 1.0$, $\rho_L =
1.0$, $v_L = 0.9$ and $p_R = 10.0$, $\rho_R = 1.0$, $v_R = 0.0$, where
$L$ stands for the left state, and $R$ the right state.  The fluid is
assumed to be an ideal gas with an adiabatic index $\Gamma = 4/3$.
The initial discontinuity is at $x = 0.5$.  The results are shown in
Fig.~\ref{fig:rie1d3}.  In this problem a strong reverse shock forms
in which post-shock oscillations are visible for the U-PPM and U-PLM
simulations, especially in the pressure profile.  In the F-WENO
simulation, the post-shock pressure oscillations is smaller those in
the U-X simulations.  In the F-PLM simulation the post-shock pressure
oscillation is nearly invisible to the eye, though they are still
present at the 0.1\% level. Reducing the CFL number from 0.5
decreases the post-shock oscillations but does not eliminate them
completely.  Table~\ref{tab:rie1d3} presents the $L_1$ norm errors and
$L_1$ order convergence rates for this problem.  Note that the
U-PPM scheme is more accurate and converges faster for this problem
than other schemes.  This is due to the fact that U-PPM scheme
captures the contact discontinuity in fewer zones (\ref{fig:rie1d3}).

% f3

% tab 3
\begin{deluxetable}{cccc}
%\tabletypesize{\scriptsize} 
\tablecaption{$L_1$ errors of the density for the 1D Riemann Problem
  3.  Four schemes with various resolutions are shown at $t=0.4$.
\label{tab:rie1d3}}
\tablewidth{0pt}
\tablehead{
\colhead{Scheme}    &   \colhead{$N$\tablenotemark{a}}  &  
\colhead{$L_1$ Error} & \colhead{Convergence Rate}
}
\startdata
F-WENO & 100  & 9.97e-2 &      \\ 
       & 200  & 6.29e-2 & 0.67 \\ 
       & 400  & 3.01e-2 & 1.1  \\ 
       & 800  & 1.69e-2 & 0.83 \\
       & 1600 & 9.48e-3 & 0.83 \\ 
       & 3200 & 5.24e-3 & 0.86 \\ 
\tableline
F-PLM  & 100  & 1.12e-1 &      \\ 
       & 200  & 6.98e-2 & 0.68  \\ 
       & 400  & 3.45e-2 & 1.0  \\ 
       & 800  & 1.94e-2 & 0.83 \\ 
       & 1600 & 1.13e-2 & 0.78 \\ 
       & 3200 & 6.54e-3 & 0.79 \\ 
\tableline
U-PPM  & 100  & 9.72e-2 &      \\ 
       & 200  & 5.60e-2 & 0.80 \\ 
       & 400  & 2.49e-2 & 1.2  \\ 
       & 800  & 1.30e-2 & 0.94 \\ 
       & 1600 & 6.06e-3 & 1.1  \\ 
       & 3200 & 3.11e-3 & 0.96 \\ 
\tableline
U-PLM  & 100  & 9.53e-2 &      \\ 
       & 200  & 6.32e-2 & 0.59 \\ 
       & 400  & 2.99e-2 & 1.1  \\ 
       & 800  & 1.78e-2 & 0.75 \\ 
       & 1600 & 1.04e-2 & 0.78 \\ 
       & 3200 & 6.10e-3 & 0.77 
\enddata
\tablenotetext{a}{Number of grid points}
\end{deluxetable}

\subsection{One-Dimensional Riemann Problem With Non-Zero Transverse
  Velocity: Easy Test}
\label{sec:rievy1}

Many problems of interest in hydrodynamics involve strong shear flows.
Astrophysical jets include shearing layers where mixing of ambient
material into the fast jet flow may be important.  It is therefore
important to test the ability of numerical codes to handle Riemann
problems with velocity components transverse to the direction of
propagation of the main flow. In Newtonian hydrodynamics the
transverse momentum is simply advected with the flow and is not
coupled directly to the equation of motion in the longitudinal
direction.  No serious difficulty is introduced by the presence of
transverse velocity components though transverse kinetic energy
dissipated through viscous dissipation can alter the longitudinal
motion by affecting the pressure. However, in relativistic flow,
transverse velocities are directly coupled to the dynamics along all
directions by the Lorentz factor which depends on all velocity
components.  This coupling makes relativistic Riemann problems with
transverse velocity much more difficult to solve correctly in a
numerical code.  In particular, higher resolution is needed.
Under-resolved simulations produce incorrect shock positions along the
normal direction.  Here, we present relatively easy 1D tests of
Riemann problems including transverse velocity which can be resolved
with moderate resolution.  In \S~\ref{sec:rievy2} we show tests
requiring high resolution and demonstrate the ability of adaptive mesh
refinement to accurately simulate very relativistic flows with
significant transverse velocity components.

\begin{figure}
%\epsscale{1.0}
%\plotone{rievy1.eps}
\plotone{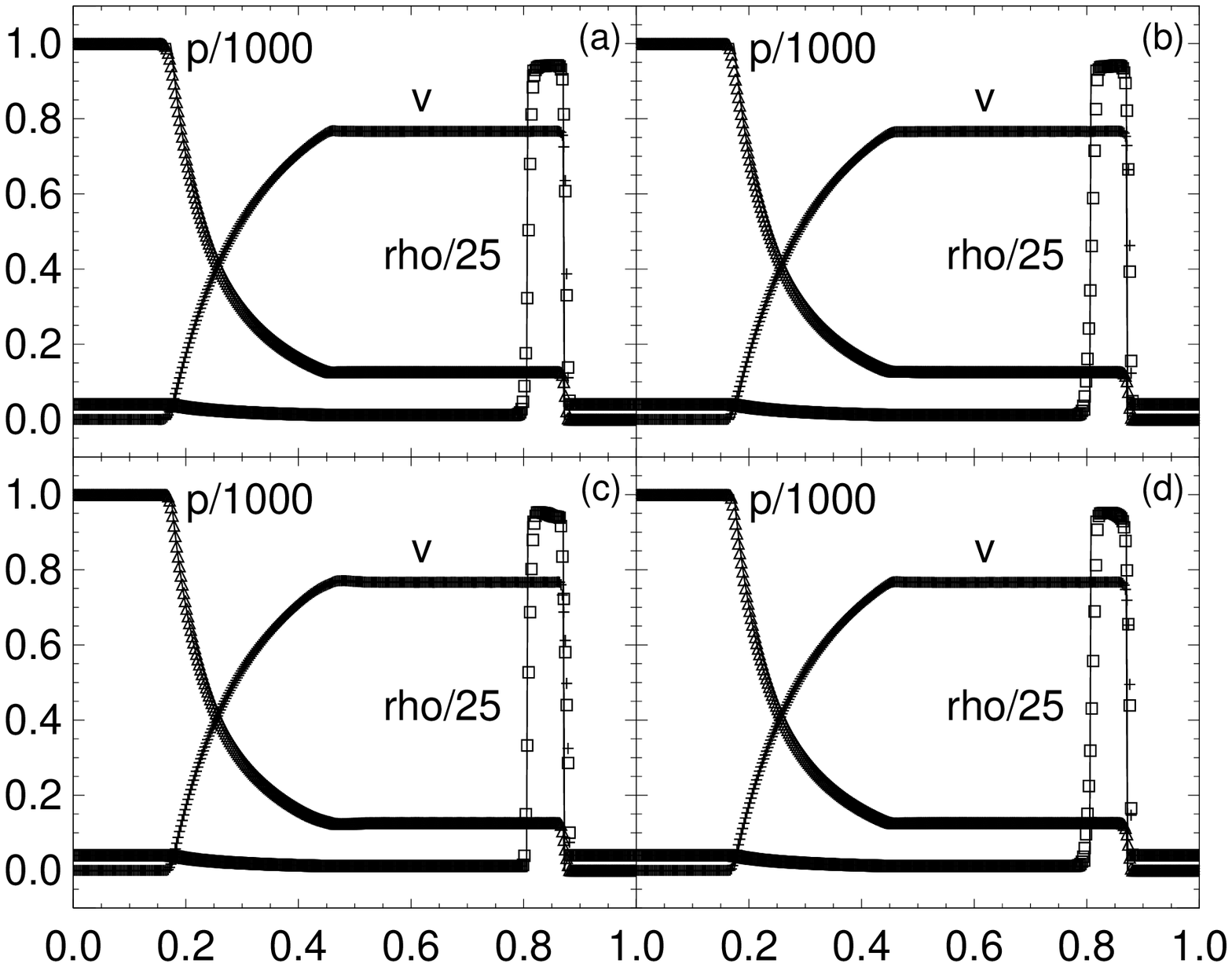}
\caption{``Easy'' one-dimensional Riemann problem with non-zero
  transverse velocity ($v_y = 0.99 c$) at $t=0.4$.  Results for four
  schemes: (a) F-WENO; (b) F-PLM; (c) U-PPM; (d) U-PLM are shown.  The
  computational grid consists of 400 zones.  Numerical results are
  shown in symbols, whereas the exact solution is shown in solid
  lines.  We show proper mass density (square), pressure (triangle)
  and velocity in $x$-direction (plus sign).
\label{fig:rievy1}}
\end{figure}

In this test, the one-dimensional numerical region ($0 \le x \le 1$)
initially consists of two constant states: $p_L = 1000.0$, $\rho_L =
1.0$, $v_{xL} = 0.0$, $v_{yL} = 0.0$ and $p_R = 10^{-2}$, $\rho_R =
1.0$, $v_{xR} = 0.0$, $v_{yR} = 0.99$, where $L$ stands for the left
state, and $R$ the right state.  The fluid is assumed to be an ideal
gas with an adiabatic index $\Gamma = 5/3$.  The initial discontinuity
is at $x = 0.5$.  The results are shown in Fig.~\ref{fig:rievy1}.  In
this test, the presence of transverse velocity alters the structure of
the Riemann problem \citep{exact,rz02}.  This problem is relatively
easy because the transverse velocity is in the cold gas of the right
state, not in the relativistically hot left state or in the
rarefaction fan which subsequently propagates into it.  The simulation
agrees well with the analytic solution even at the relatively modest
resolution of 400 zones.  Table~\ref{tab:rievy1} presents the $L_1$
norm errors and $L_1$ order convergence rates for this problem.

% f4

% tab 4
\begin{deluxetable}{cccc}
%\tabletypesize{\scriptsize} 
\tablecaption{$L_1$ errors of the density for the ``easy'' 1D Riemann
  problem with non-zero transverse velocity at $t=0.4$.  Four schemes
  with various resolutions using a uniform grid are shown.
\label{tab:rievy1}}
\tablewidth{0pt}
\tablehead{
\colhead{Scheme}    &   \colhead{$N$\tablenotemark{a}}  &  
\colhead{$L_1$ Error} & \colhead{Convergence Rate}
}
\startdata
F-WENO & 100  & 7.58e-1 &      \\ 
       & 200  & 3.92e-1 & 0.95 \\ 
       & 400  & 2.31e-1 & 0.76 \\ 
       & 800  & 1.18e-1 & 0.97 \\
       & 1600 & 6.58e-2 & 0.84 \\ 
       & 3200 & 3.44e-2 & 0.94 \\ 
\tableline
F-PLM  & 100  & 8.26e-1 &      \\ 
       & 200  & 4.59e-1 & 0.85  \\ 
       & 400  & 2.77e-1 & 0.73  \\ 
       & 800  & 1.49e-1 & 0.89 \\ 
       & 1600 & 8.00e-2 & 0.90 \\ 
       & 3200 & 4.63e-2 & 0.79 \\ 
\tableline
U-PPM  & 100  & 8.48e-1 &      \\ 
       & 200  & 4.25e-1 & 1.0  \\ 
       & 400  & 2.41e-1 & 0.82  \\ 
       & 800  & 1.27e-1 & 0.92 \\ 
       & 1600 & 6.43e-2 & 0.99  \\ 
       & 3200 & 3.34e-2 & 0.95 \\ 
\tableline
U-PLM  & 100  & 9.00e-1 &      \\ 
       & 200  & 4.72e-1 & 0.93 \\ 
       & 400  & 2.88e-1 & 0.71  \\ 
       & 800  & 1.52e-1 & 0.92 \\ 
       & 1600 & 8.86e-2 & 0.78 \\ 
       & 3200 & 4.95e-2 & 0.84 
\enddata
\tablenotetext{a}{Number of grid points}
\end{deluxetable}

\subsection{One-Dimensional Shock Heating Problem In Planar Geometry}

\begin{figure}
%\epsscale{1.0}
%\plotone{heating1d.eps}
\plotone{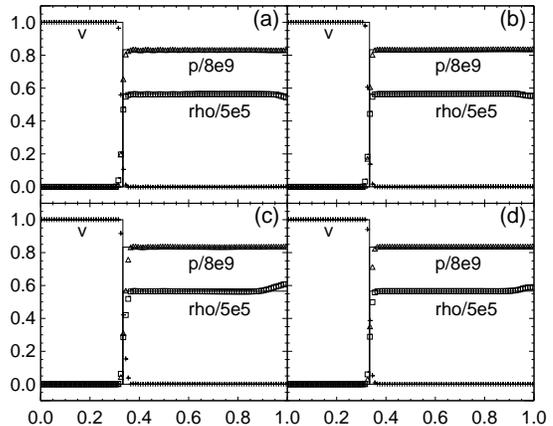}
\caption{ One-dimensional shock heating problem in planar geometry at
  $t=2.0$.  Results for four schemes: (a) F-WENO; (b) F-PLM;
  (c) U-PPM; (d) U-PLM are shown.  The parameter $\theta$ in the
  minmod slope limiter for U-PLM is set to be $1.0$ in this test.  The
  computational grid consists of 100 zones.  Numerical results are
  shown in symbols, whereas the exact solution is shown in solid
  lines.  We show proper mass density (square), pressure (triangle)
  and velocity (plus sign).
\label{fig:heating1d}}
\end{figure}

The shock heating problem is a standard test to study the ability of a
code to handle very strong shocks with sufficiently few zones and
without excessive post shock oscillations.  Cold gas flows toward a
reflecting boundary at $x=1.0$ and a reverse strong shock forms and
propagates to the left decelerating the gas to zero speed.  The gas
has an initial speed of $v = 0.9999999999 = 1.0 - 10^{-10}$
(corresponding to a Lorentz factor of $W = 70710.675$) and initial
density of $\rho = 1.0$.  Due to conservation of energy and the fact
the gas initially has nearly zero internal energy compared to kinetic
energy (we use a small value $\epsilon_0 = 0.003$ for numerical
reasons), the specific internal energy after the shock is simply
$\epsilon = W - 1$.  The compression ratio across the shock can be
arbitrarily large in the relativistic case and is given by
\begin{equation}
\sigma = \frac{\Gamma+1}{\Gamma-1} + \frac{\Gamma}{\Gamma-1} \epsilon
\label{sigma}
\end{equation}
where $\Gamma $ is the adiabatic index of the gas.  In this case
$\Gamma = 4/3$ so $\sigma = 282845.70$ and the shock velocity is given
by $v_s = \frac{(\Gamma-1)W|v|}{W+1} = 0.33332862$.  In
Fig.~\ref{fig:heating1d} we show our results for a uniform mesh of 100
zones compared with the analytic solution at $t=2.0$.

% f5

We note that for this problem with a constant state behind the shock,
as opposed to a thin shell which might more naturally occur in a
realistic flow, the maximum Lorentz factor that can be achieved
numerically is just limited by floating point precision.

The parameter $\theta$ in the minmod slope limiter for U-PLM is set to
be $1.0$ in this test to eliminate strong oscillations which would
appear if the default value $\theta = 1.5$ is used in U-PLM.  In
this test with U-PPM, stronger dissipation is also needed to avoid
crashes.  In particular, one of the PPM parameters, $\epsilon^{(2)}$,
is set to be $10^{-4}$ instead of the default value $\epsilon^{(2)} =
1.0$ \citep[see][for the meaning of this parameter]{ppm}.  The
reflecting wall at $x=1.0$ poses difficulties for numerical
simulations and gives rise to visible errors for zones near the wall.
The errors of density at the nearest zone to the reflecting wall are
3.9\%, 2.4\%, 7.4\%, and 4.3\%, for F-WENO, F-PLM, U-PPM, and U-PLM,
respectively.  For this particular problem with very strong shock but
simple structure, more diffusive schemes F-PLM and U-PLM perform
better than F-WENO and U-PPM.

\subsection{Isentropic Smooth Flows}
\label{sec:isen}

Most tests presented in this paper involve strong shocks.  In addition
to tests with discontinuities, it is also very important to test the
capability of the schemes to handle smooth flows.  In this section, we
present two tests involving the isentropic evolution of smooth flows
in 1D and 2D. 

\begin{figure}
\epsscale{0.75}
%\plotone{isen.eps}
\plotone{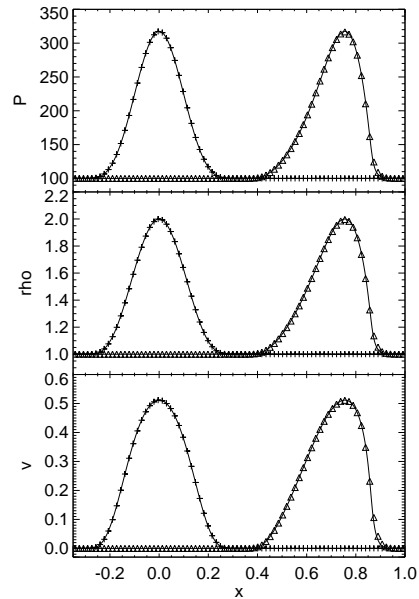}
\caption{ One-dimensional isentropic flow.  The initial pulse is shown
  on the left, and the pulse at $t = 0.8$ on the right.  Pressure,
  density and velocity are shown in the top, middle and bottom panels
  respectively.  The solid lines are exact solutions.  Numerical
  results of F-WENO at $t = 0$ (plus signs) and $t = 0.8$ (triangle)
  are shown.
\label{fig:isen}}
\end{figure}

In the first test, a one-dimensional isentropic smooth pulse is set up
in a uniform reference state, and the run stops some time before a
shock is formed.  The test is similar to the convergence tests
performed by \citet{col06} for their Newtonian hydrodynamics code.
The initial density structure at $t = 0$ is given by
\begin{equation}
  \rho_0(x) = \rho_{\mathrm{ref}} (1 + \alpha f(x)),
\label{eq:rhocolella}
\end{equation}
where $\rho_{\mathrm{ref}}$ is the density of the reference state and
\begin{equation}
  f(x) = \left\{ \begin{array}
                  {l@{\quad:\quad}l}
                 ((x/L)^2 - 1)^4 & |x|<L  \\ 0 & \mathrm{otherwise},
                  \end{array} \right.
\label{eq:fcolella}
\end{equation}
here $\alpha$ is the amplitude of the pulse and $L$ the width of the
pulse.  The pressure is given by the isentropic relation, $p = K
\rho^\Gamma$, where $K$ is a constant.  The initial velocity of the
reference state is $v_{\mathrm{ref}} = 0$.  The initial velocity inside
the pulse at $t = 0$ is set up by assuming one of the two Riemann
invariants,
\begin{equation}
  J_- = \frac{1}{2} \ln (\frac{1+v}{1-v}) - 
  \frac{1}{\sqrt{\Gamma-1}}
  \ln (\frac{\sqrt{\Gamma-1}+c_s}{\sqrt{\Gamma-1}-c_s} ),
\end{equation} 
is constant across the whole region, where $c_s$ is the sound speed.
We note that the other Riemann invariant,
\begin{equation}
  J_+ = \frac{1}{2} \ln (\frac{1+v}{1-v}) + 
  \frac{1}{\sqrt{\Gamma-1}}
  \ln (\frac{\sqrt{\Gamma-1}+c_s}{\sqrt{\Gamma-1}-c_s} )
\end{equation} 
is not constant.  The exact solution of the test can be obtained by
using standard characteristic analysis.  The pulse will have a smooth
shape before a shock eventually forms.  The width and height of the
pulse does not change before the shock forms.  But the pulse will
become increasingly asymmetric as the shape of the front of the pulse
becomes steeper during the propagation (see Fig.~\ref{fig:isen}).
Behind the moving pulse, the fluid goes back to the reference state.

% f6

Our computational region for this test is $-0.35 \le x \le 1$.  The
reference state is, $p_\mathrm{ref} = 100, \rho_\mathrm{ref}=1,
v_\mathrm{ref} = 0$, the amplitude of the pulse is $\alpha = 1.0$, and
the width is $L = 0.3$.  The adiabatic index in the equation of state
is $\Gamma = 5/3$.

We have run this test with different schemes and various numerical
resolution.  Because the flow is very smooth, all methods perform very
well for this test.  Fig.~\ref{fig:isen} shows the exact solution and
the numerical results of F-WENO with 80 uniform zones.  The results of
the convergence rates are shown in Table~\ref{tab:isen}.  Various
Runge-Kutta methods, including the third-order TVD scheme, standard
fourth-order scheme (RK4) and fifth-order scheme (RK5) have been used
in this test.  We find that the F-WENO scheme, which is fifth-order
accurate in space and third-order accurate in time, has an $L_1$ order
convergence rate of $\sim 3-4$ for this one-dimensional test with
smooth flow.  For the F-WENO-RK4 and F-WENO-RK5 schemes, which are
fourth and fifth-order accurate in time integration respectively,
order of convergence rates up to $\sim 5$ can be achieved.  Note that
the difference between the results of RK4 and RK5 is very small.  For
schemes other than WENO, including schemes using RK4 or RK5, the order
of convergence rate is about 2 (Table~\ref{tab:isen}).  These results
clearly show that the fifth-order WENO scheme is very accurate and
converges very quickly for smooth flows.  

% tab 4.5
\begin{deluxetable}{cccc}
%\tabletypesize{\scriptsize} 
\tablecaption{$L_1$ errors of the density for the 1D isentropic flow
  problem at $t=0.8$.  Results with various resolutions using a
  uniform grid are shown.
\label{tab:isen}}
\tablewidth{0pt}
\tablehead{
\colhead{Scheme\tablenotemark{a}}   &   \colhead{$N$\tablenotemark{b}}  &  
\colhead{$L_1$ Error} & \colhead{Convergence Rate}}
\startdata
F-WENO  &    80 &  2.07e-3    &    \\
        &   160 &  1.10e-4    &     4.2  \\
        &   320 &  1.70e-5  &     2.7  \\
        &   640 &   1.47e-6  &     3.5  \\
       	&  1280 &   1.58e-7  &     3.2  \\
       	&  2560 &   1.91e-8  &     3.1  \\
       	&  5120 &   2.37e-9  &     3.0  \\      
\tableline
F-WENO-RK4 &   80 &  1.87e-3   &   \\
	   &  160 &  1.18e-4   &    4.0  \\
	   &  320 &   1.31e-5   &    3.2  \\
	 &    640 &   6.80e-7   &    4.3  \\
	 &   1280 &   2.54e-8   &    4.7  \\
	 &   2560 &   7.91e-10   &    5.0  \\
	 &   5120 &   2.38e-11   &    5.1  \\   
\tableline
F-WENO-RK5 &   80 &  1.87e-3  &    \\
  	 &   160 &   1.17e-4   &    4.0  \\
 	 &   320 &   1.30e-5   &    3.2  \\
	 &   640 &   6.82e-7   &    4.3  \\
	 &  1280 &   2.54e-8   &    4.7  \\
	 &  2560 &   8.01e-10   &    5.0  \\
	 &  5120 &   2.40e-11   &    5.1  \\    
\tableline
F-PLM	 &    80 &   8.79e-3   &   \\
         &   160 &   4.05e-3   &    1.1  \\
         &   320 &   1.22e-3   &    1.7  \\
         &   640 &   3.10e-4   &    2.0  \\
	 &  1280 &   7.83e-5   &    2.0  \\
       	 &  2560 &   1.96e-5   &    2.0  \\
       	 &  5120 &   4.92e-6   &    2.0   \\   
\tableline
F-PLM-RK4 &   80 &   8.85e-3  &    \\
	  &  160 &   4.06e-3   &    1.1  \\
 	  &  320 &   1.21e-3   &    1.7  \\
 	  &  640 &   3.11e-4   &    2.0  \\
	  &  1280 &  7.84e-5   &   2.0 \\ 
	 &  2560 &   1.97e-5   &   2.0 \\
	&   5120 &   4.93e-6   &   2.0 \\ 
\tableline
U-PPM &	     80 &    1.11e-2   &   \\
 	&    160 &   2.47e-3   &    2.2 \\ 
 	&    320 &   7.02e-4   &    1.8 \\
 	&    640 &   1.38e-4   &    2.3 \\ 
	&   1280 &   2.92e-5   &    2.2 \\ 
	&   2560 &   6.48e-6   &    2.2 \\
	&   5120 &   1.52e-6   &    2.1 \\
\tableline
U-PPM-RK4 &   80 &   1.10e-2  &    \\
 	  &  160 &   2.56e-3  &     2.1  \\
 	  &  320 &   5.74e-4  &     2.2  \\
 	  &  640 &   1.34e-4   &    2.1  \\
	  &  1280 &  3.10e-5  &     2.1  \\
	  &  2560 &  7.33e-6  &     2.1  \\
	 &  5120 &   1.82e-6   &    2.1  \\
\tableline
U-PLM    &    80 &   1.12e-2  &      \\
         &   160 &   3.56e-3  &     1.7  \\
         &   320 &   1.03e-3  &     1.8  \\
         &   640 &   2.61e-4  &     2.0 \\ 
       	 &  1280 &   6.50e-5  &     2.0  \\ 
       	 &  2560 &   1.62e-5  &     2.0  \\ 
       	 &  5120 &   4.03e-6  &     2.0  \\
\tableline
U-PLM-RK4 &   80 &    1.12e-2   &    \\
 	  &   160 &   3.56e-3   &     1.7  \\
 	  &   320 &   1.02e-3   &    1.8  \\
 	  &   640 &   2.60e-4   &    2.0  \\
	  &  1280 &   6.49e-5   &    2.0  \\
	  &  2560 &   1.62e-5   &    2.0  \\
	  &  5120 &   4.04e-6   &    2.0  
\enddata
\tablenotetext{a}{RK4 and RK5 denote the fourth and fifth-order
  Runge-Kutta methods, respectively.  The third-order Runge-Kutta
  method (RK3) is used unless otherwise stated.}
\tablenotetext{b}{Number of grid points}
\end{deluxetable}

PPM is generally 3rd-order accurate in space, but the flattening
procedure and the requirement of monotonic profiles degrades the
accuracy of the scheme at places like local extrema or where the
second derivatives of variables are large.  Thus its accuracy can be
as low as first-order in some places.  Moreover, in the U-PPM scheme,
the reconstruction is carried out on primitive variables, not
conservative variables.  The conservative variables are cell averages.
But the primitive variables converted from conservative variables are
not exactly cell averages.  However, the PPM reconstruction algorithm
regards them as cell averages.  Therefore the interface values of
primitive variables may not have third-order accuracy.  Thus it is
reasonable that the U-PPM scheme does not achieve a third-order
convergence rate.

In the second test, we have performed two-dimensional calculations to
assess the convergence rate of the WENO scheme in multi-dimension.
The computational region in this test consists of a two-dimensional
box in Cartesian coordinates with $0.0 \le x \le 3.75$ and $0.0 \le y
\le 5.0$.  The boundary conditions are periodic for all four sides of
the box.  Like the one-dimensional test of isentropic flows, there is
a static uniform reference state, which is set to $p_\mathrm{ref} =
100, \rho_\mathrm{ref}=1, v_\mathrm{ref} = 0$.  Pulses which are
periodic in space are added to the system.  Along the direction of
$\mathbf{k} = (4/5,3/5)$, the profile is periodic with a spatial period
of $S=3.0$, and the profile is constant along the direction
perpendicular to the vector $\mathbf{k}$.  Thus the spatial periods
along the $x$ and $y$-direction are 3.75 and 5.0, respectively.  Note
that these are consistent with the size of the computational box with
periodic boundaries.  The pulses move along the direction of the
vector $\mathbf{k}$.  The initial density profile is given by
$\rho_0(d)$ (Eqs.~\ref{eq:rhocolella} \& \ref{eq:fcolella}), where
$d$, the distance to the center of the nearest pulse, is given by $d =
\mathrm{mod}(\mathbf{k} \cdot \mathbf{r} + S/2, S) - S/2$, here
$\mathrm{mod}(a,b)$ returns the reminder of the division $a/b$, and
$\mathbf{r} = (x,y)$.  The amplitude of the pulse is $\alpha = 1.0$,
and the width is $L = 0.9$.  The adiabatic index in the equation of
state is $\Gamma = 5/3$.  Similar to the one-dimensional test, the
initial pressure is given by the isentropic relation, and the initial
velocity by assuming the Riemann invariant, $J_-$ is constant.

We have run the two-dimensional test using the WENO scheme with three
Runge-Kutta methods: RK3, RK4 and RK5.  Results of the $L_1$ norm
errors and convergence rate are shown in Table~\ref{tab:isen2d}.
Similar to the one-dimensional test, both F-WENO-RK4 and F-WENO-RK5
performs better than F-WENO, which uses RK3.  But the difference of
errors between F-WENO-RK4 and F-WENO-RK5 is very small.  Therefore,
for this test it is not worth using the more expensive RK5 for time
integration.

% tab 4.75
\begin{deluxetable}{cccc}
%\tabletypesize{\scriptsize} 
\tablecaption{$L_1$ errors of the density for the 2D isentropic flow
  problem at $t=2.4$.  Results with various resolutions using a
  uniform grid are shown.
\label{tab:isen2d}}
\tablewidth{0pt}
\tablehead{
\colhead{Scheme\tablenotemark{a}}   &   \colhead{$N$\tablenotemark{b}}  &  
\colhead{$L_1$ Error} & \colhead{Convergence Rate}}
\startdata
F-WENO  &  $48 \times 64 $    &  7.35e-2  &      \\
        &  $96 \times 128 $   &  4.43e-3  &  4.1 \\
        &  $192 \times 256 $  &  8.04e-4  &  2.5 \\
        &  $384 \times 512 $  &  9.62e-5  &  3.1 \\
        &  $768 \times 1024 $ &  1.12e-5  &  3.1 \\ 
\tableline
F-WENO-RK4  &  $48 \times 64 $    &  7.24e-2  &      \\
            &  $96 \times 128 $   &  4.75e-3  &  3.9 \\
            &  $192 \times 256 $  &  4.70e-4  &  3.3 \\
            &  $384 \times 512 $  &  3.18e-5  &  3.9 \\
            &  $768 \times 1024 $ &  1.24e-6  &  4.7 \\
\tableline
F-WENO-RK5  &  $48 \times 64 $    &  7.19e-2  &      \\
            &  $96 \times 128 $   &  4.67e-3  &  3.9 \\
            &  $192 \times 256 $  &  4.61e-4  &  3.3 \\
            &  $384 \times 512 $  &  3.13e-5  &  3.9 \\
            &  $768 \times 1024 $ &  1.22e-6  &  4.7  
\enddata
\tablenotetext{a}{The third, fourth (RK4), and fifth-order (RK5)
  Runge-Kutta methods are used with the fifth-order WENO scheme.}
\tablenotetext{b}{Number of grid points in $x$ and $y$-direction}
\end{deluxetable}

\subsection{Two-Dimensional Tests: Wind Tunnel With Step}
\label{sec:emery}

In order to test the ability of our code to handle strong shocks in
multiple dimensions, we have performed standard tests published in the
literature.

The Emery step \citep{emery,wc84} consists of a horizontal wind flowing
into a step, represented as a reflecting boundary condition.  The
corner of the step represents a singular point of the rarefaction fan.
As the wind collides with the step a reverse shock propagates back
into the wind forming a bow shock which reflects off the upper
boundary and forms a Mach stem which should remain straight and
initially almost vertical.  Since this structure can remain aligned
with the vertical coordinate of the numerical grid, problems such as
``odd-even decoupling'' in some numerical schemes can act to
incorrectly modify the stem \citep{quirk}.

We simulate the Emery step in a computational box with $0 \le x \le 3$
and $0 \le y \le 1$ with 240 zones in the $x$-direction and 80 in the
$y$-direction.  The step is represented as a reflecting boundary of
height 0.2 beginning at $x= 0.6$ and continuing along the length of
the box.  The upper boundary and lower boundary for $x<0.6$ are also
both reflecting.  Initially the box is filled with $\rho =1.4$ gas
moving at $v_x = 0.999$.  The left boundary is an inflow with the same
quantities.  The right boundary is outflow.  The adiabatic index in
the equation of state is $\Gamma = 1.4$.  The Newtonian Mach number is
3.0, and the corresponding relativistic Mach number $\mathcal{M} = W M
/ W_s$, where $M$ is the classical Mach number, $W$ is the Lorentz
factor of the gas and $W_s$ is the Lorentz factor of sound speed in
the gas, is $\sim 63$.

\begin{figure}
%\epsscale{1.0}
%\plotone{emery.eps}
\plotone{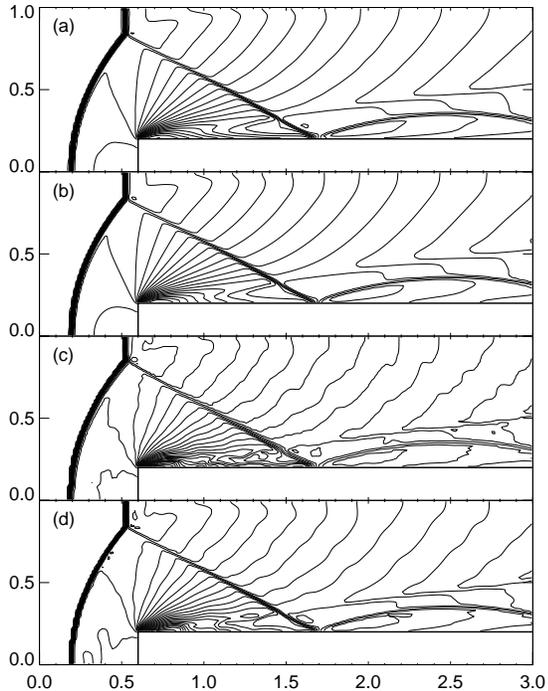}
\caption{ Emery step at $t=4.0$ with $240 \times 80$ resolution.
  Thirty equally spaced contours of the logarithm of proper density
  are plotted.  Results for four schemes: (a) F-WENO; (b) F-PLM; (c)
  U-PPM; (d) U-PLM are shown.
\label{fig:emery}}
\end{figure}

Our results are shown in Fig.~\ref{fig:emery}.  These results are
comparable to those of \citet{ls04}.  Some numerical methods require
special entropy fixes to accurately simulate the flow near the step
corner.  No such code modification was implemented or required for the
simulations presented here because we are only interested in the
global structures, especially the shocks.  Though numerical boundary
artifacts exist near the corner, the global solutions are not
affected.

% f7

The F-PLM and F-WENO methods are clearly superior for this problem.
U-PPM and U-PLM are noisy both downstream of the reverse shock and
along the top of the step.  Presumably the results of U-PPM and U-PLM
could be improved by tuning the PPM parameters and the $\theta$
parameter in the minmod slope limiter.  For this problem we use the
standard CFL number of 0.5.  Reducing this number could not improve
the results for U-PPM and U-PLM.

A high resolution run employing AMR is shown in
\S~\ref{sec:emery_amr}.

\subsection{Two Dimensional Tests: Shock Tube}
\label{sec:rie2d}

In order to compare with existing multi-dimensional SRHD codes we
repeat the two-dimensional shock tube problem suggested by
\citet{db02} and repeated by \citet{ls04}.  This test consists of a
Cartesian box divided into four equal area constant states:
$$ \begin{array}{ll}
(\rho,v_x,v_y,p)^{NE} =  & (0.1,0,0,0.01), \\
(\rho,v_x,v_y,p)^{NW} =  & (0.1,0.99,0,1), \\
(\rho,v_x,v_y,p)^{SW} =  & (0.5,0,0,1),  \\
(\rho,v_x,v_y,p)^{SE} =  & (0.1,0,0.99,1), 
\end{array} $$
where NE stands for the upper right quarter of the box ($0.5 \leq x
\leq 1.0 $, $0.5 \leq y \leq 1.0 $), NW the upper left quarter ($0.0
\leq x < 0.5 $, $0.5 \leq y \leq 1.0 $), SW the lower left quarter
($0.0 \leq x < 0.5 $, $0.0 \leq y < 0.5 $) and SE the lower right
quarter ($0.5 \leq x \leq 1.0 $, $0.0 \leq y < 0.5 $).  We use
constant zoning of $400 \times 400$ zones, a $\Gamma = 5/3$ adiabatic
equation of state, outflow boundary conditions in all directions and a
CFL number of 0.5.  

In the F-PLM run of this test, the parameter $\theta$ in the minmod
slope limiter is set to $1.2$ instead of the default value $1.5$ to
avoid crashes.

\begin{figure}
%\epsscale{1.0}
%\plotone{riem2d.eps}
\plotone{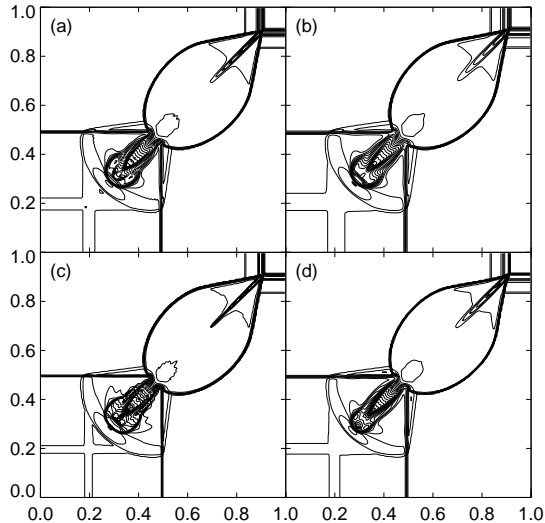}
\caption{ Two dimensional shock tube problem at $t = 0.4$.  A $\Gamma
  = 5/3$ adiabatic equation of state was used with outflow boundary
  conditions at all boundaries, $400 \times 400$ Cartesian zones and a
  CFL number of 0.5.  Results of F-WENO, F-PLM, U-PPM, and U-PLM are
  shown in panels (a), (b), (c), and (d), respectively.  Thirty
  equally spaced contours of the logarithm of proper density are
  plotted.
\label{fig:riemann2d}}
\end{figure}

Our results of four schemes are shown in Fig.~\ref{fig:riemann2d}.
The interface between panels NW and SW and the interface between
panels SE and SW are stationary contact discontinuities with jumps in
transverse velocity.  We note that the contact discontinuities have
been smeared and numerical artifacts are clearly shown in the density
contours at panel SW, as in the results in the literature
\citep{db02,ls04}.  Indeed less prominent numerical artifacts also
exist in panels NW and SE, though they do not show up in the figures
of 30 contours.  The appearance of two curved shocks and the elongated
diagonal spike of density in between in panel NE is in agreement with
the results of \citet{ls04}, whereas the diagonal feature is much less
prominent in the results of \citet{db02}.  We note that the initial
interface between panels NE and NW and the interface between panels NE
and SE are not simple shock waves.  A similar test problem with simple
waves at all initial interfaces (two contact discontinuities and two
shocks) has been performed by \citet{mig05} to test their
multidimensional relativistic PPM code.

% f8

\section{Adaptive Mesh Refinement}
\label{sec:amr}

Solutions to hyperbolic partial differential equations (PDEs) are
frequently smooth in large fractions of the computational volume yet
contain sharp transitions in localized regions.  In the smooth
regions, relatively coarse numerical zoning may be sufficient to
accurately represent the solution, while finer zoning is needed where
sharp transitions occur.  Adaptive mesh refinement \citep{bo84,bc89}
creates finer numerical meshes to adequately resolve steep gradients
and thin layers where they exist.  For computational efficiency,
coarser meshes are used in smooth regions so that computing power is
not wasted over-resolving regions where high resolution is
unnecessary.  As the solution evolves, the mesh structure adapts to
provide appropriate resolution where needed at a given time.

The SRHD equations are PDEs which admit the formation of extremely sharp
transitions in the form of shock waves, thin shells and contact
discontinuities.  These structures coexist with smoothly flowing regions in
different parts of the computational domain.  AMR is a powerful tool to
handle the resolution required to adequately capture these features.

In the RAM code, we utilize the AMR tools in the FLASH code version
2.3 \citep{flash}, which in turn is a modified version of the PARAMESH
AMR package \citep{paramesh}.  PARAMESH is a block-structured AMR
package.  It uses a hierarchy of nested, logically Cartesian blocks
which typically have eight zones per dimension for a total of $8^d$
zones per block, where $d=$ 1, 2 or 3 is the dimensionality of the
simulation.  Finer level blocks are a factor of two higher in
resolution in each direction so that each block is either at the
highest level of refinement or contains $2^d$ daughter blocks.  Flux
conservation at jumps of refinement is imposed by replacing fluxes
computed at the courser level of refinement with appropriate sums of
fluxes at the finer level.

Refinement or derefinement of a block is generally determined by
calculating an approximate numerical second derivative of fluid
variables which can be specified at runtime.  Other kinds of criteria
can also be used for specific problems (see e.g., \S~\ref{sec:kh}).
In FLASH, the one-dimensional normalized second derivative, which is a
measurement of error, is given by\footnote{Note that the expression
  for the error norm in FLASH~2.3, on which our RAM code is based, is
  slightly different from that in \citet{flash}},
\begin{equation}
  E_i = \frac{|u_{i+2} - 2 u_i + u_{i-2}|}{|u_{i+2}-u_i| +
    |u_i-u_{i-2}| + \epsilon (|u_{i+2}| + 2|u_i| + |u_{i-2}|)},
\end{equation}
where the last term in the denominator is a low pass filter to avoid
excessive refinement on small fluctuations, and $\epsilon$ is an
adjustable parameter for the low-pass filter.  By default, we use
$\epsilon = 0.01$ except for the test problem in \S~\ref{sec:rievy2}.
The one-dimensional expression can be generalized to multi-dimension
\citep{flash}.  Typically pressure, density and Lorentz factor are
used to estimate the error norm, $E$, once every two steps.  If the
maximum error norm on a block is larger than the value of a parameter,
$E_{\mathrm{ref}}$, the block will be marked for refinement.  If the
maximum error norm on a block is less than the value of a parameter,
$E_{\mathrm{deref}}$, the block will be marked for derefinement.  The
default values for the above two parameters in the tests shown in this
paper are set to $E_{\mathrm{ref}} = 0.8$, and $E_{\mathrm{deref}} =
0.2$ except for the test problem in \S~\ref{sec:rievy2}.  When AMR
refines a block, quadratic interpolation algorithms are used for
prolongation operations between refinement levels.  The prolongation
operation is performed on the conserved variables.  Then the
conversion of conserved variables to primitive variables is performed.
The conversion could fail in principle, but this has never happened in
our simulations because the refinement is performed when the solution
is still very smooth.  If the conversion fails, the problem can be
fixed by using first-order prolongation.  The solutions on parent
blocks, which are not evolved, are obtained by restriction operations.

In \S~\ref{sec:fallback}, we discussed the fall-back mechanism, which
makes our code more robust.  When AMR is used, the cells in which
unphysical results appear are almost always at the finest level of
refinement.  Thus we only need to apply the more diffusive schemes for
reconstruction to the finest level.  For coarser levels, we still use
the standard reconstruction scheme in the recalculation.

FLASH handles parallelization using the message passing interface
(MPI) library and uses an estimate of the work per processor to
balance the computational load among processors.

\section{Tests With Adaptive Mesh Refinement}
\label{sec:amrtest}

In this section, we present results of the AMR on some test problems.
All the tests shown are performed with the F-WENO scheme, which is
very robust and accurate for all the tests we have performed.  We use
F-WENO-A and F-WENO-U to denote the F-WENO scheme with adaptive mesh
and uniform grid, respectively.

\subsection{One-Dimensional Riemann Problem With Transverse Velocity:
  Hard Test}
\label{sec:rievy2}

In \S~\ref{sec:rievy1}, we tested out schemes on a one-dimensional
Riemann problem with non-zero transverse velocity.  That test problem
is relatively easy and can be resolved with modest resolution (e.g.,
400 zones).  To fully exercise the code, we perform a very severe test
requiring very high resolution to resolve the complicate structure of
the transverse velocity.

\begin{figure}
%\epsscale{1.0}
%\plotone{rievy2.eps}
\plotone{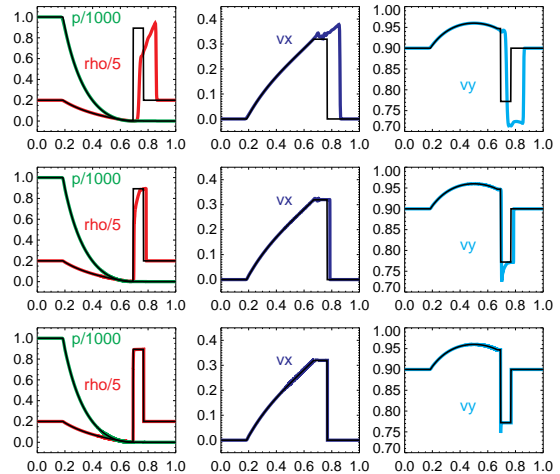}
\caption{1D Riemann problem with transverse velocity on both sides at
  $t = 0.6$.  Results of F-WENO-A with three different resolutions are
  shown.  The lowest refinement level is 1 in all three runs, while
  the highest levels are 1, 4, and 8, for top, middle, and bottom
  panels, respectively.  The level 1 contains 400 zones.  Thus the
  equivalent numbers of zones are 400, 3200, and 51200, for top,
  middle, and bottom panels, respectively.  Color lines show the
  numerical results for proper density (red), pressure (green), normal
  velocity (blue), and transverse velocity (cyan), whereas solid lines
  show the exact solutions.
\label{fig:rievy2}}
\end{figure}

In this test, the one-dimensional numerical region ($0 \le x \le 1$)
initially consists of two constant states: $p_L = 1000.0$, $\rho_L =
1.0$, $v_{xL} = 0.0$, $v_{yL} = 0.9$ and $p_R = 10^{-2}$, $\rho_R =
1.0$, $v_{xR} = 0.0$, $v_{yR} = 0.9$, where $L$ stands for the left
state, and $R$ the right state.  The fluid is assumed to be an ideal
gas with an adiabatic index $\Gamma = 5/3$.  The initial discontinuity
is at $x = 0.5$.  The results at $t = 0.6$ are shown in
Fig.~\ref{fig:rievy2}.  In this test, the breakup of the initial
discontinuity evolves into a shock moving towards the right, a rarefaction
wave moving towards the left, and a contact discontinuity in between.  It
is interesting to note the differences among this test problem, a
similar problem in \S~\ref{sec:rie1d2}, which has no transverse
velocity, and the problem in \S~\ref{sec:rievy1}, which has transverse
velocity on the right side.  The presence of large transverse velocity
on the left make the shock move slower than that in the previous
problems.  Compared to the previous problems, the density jump is
smaller and the post shock dense shell is wider.  However, these
differences do not make this problem easier.  The distance between the
tail of the rarefaction wave and the contact discontinuity is much
smaller in this test.  More interesting is the structure of
the transverse velocity.  Inside the rarefaction wave, the transverse
velocity increases from $v_y = 0.9$ at the head of the rarefaction
wave to $v_y = 0.9602$ and then decreases to $v_y=0.9472$ at the
tail of the rarefaction wave.  While the Lorentz factor increases
monotonically from $W = 2.29$ at the head to $W = 35.8$ at the tail of
the rarefaction wave.  Both the normal and transverse velocity stay
constant in the thin shell between the tail of the rarefaction wave
and the contact discontinuity.  The presence of a very large Lorentz
factor ($\sim 36$) in that thin shell on the left side of the contact
discontinuity requires extremely high resolution to resolve the
structure.  Across the contact discontinuity, pressure and normal
velocity do not change, while density and transverse velocity have a
jump in their values.  At the shock front, the post shock state has a
transverse velocity of $v_y = 0.7721$, which is smaller than the
pre-shock value $v_y = 0.9$.

% f9

The results (Fig.~\ref{fig:rievy2}) clearly demonstrate the demand of
high resolution when strong shear flows present.  In low resolution
runs, the position of the shock front and the contact discontinuity
were not correctly captured.  Even with $51200$ zones, there is still
visible errors for transverse velocity at the contact discontinuity.
In the cases like this, we argue that AMR could be a powerful tool.
For the sake of comparison, we run the test with various uniform grids
and AMR grids with correspondingly equivalent resolutions.  For this
test, the AMR parameters (\ref{sec:amr}) are set to $\epsilon =
0.005$, $E_{\mathrm{ref}} = 0.5$, and $E_{\mathrm{deref}} = 0.1$.  For
the AMR grids, the fine zoning follows the shock front and the contact
discontinuity automatically, while coarse zoning is used at smooth
region like the middle of the rarefaction wave.  In one of the runs, 8
levels of refinement are used.  That is the finest zones are 128 times
smaller than the coarsest zones.  The AMR makes the resolution of this
run equivalent to $51200$ uniform zones.  Meanwhile AMR runs are much
faster than their corresponding uniform zoning runs.  The measurement
of global errors is shown in Table~\ref{tab:rievy2}.  As we expected,
results from AMR are comparable to those from uniform zoning, even
though coarse zoning is used in most of the computational region.
This is due to the fact that the global error is dominated by the
discontinuities where high resolution is used in AMR runs.

% tab 5
\begin{deluxetable}{ccccc}
%\tabletypesize{\scriptsize} 
\tablecaption{$L_1$ errors of the density for the 1D Riemann Problem
  with transverse velocity on both sides.
\label{tab:rievy2}}
\tablewidth{0pt}
\tablehead{
\colhead{Mesh}       &
\colhead{Levels\tablenotemark{a}}  &
\colhead{$N$\tablenotemark{b}}  &  
\colhead{$L_1$ Error} & \colhead{Convergence Rate}
}
\startdata
Uniform    & 1 &    400  &  5.21e-1  &       \\
           & 1 &    800  &  3.63e-1  &  0.52 \\
           & 1 &   1600  &  2.33e-1  &  0.64 \\
           & 1 &   3200  &  1.26e-1  &  0.89 \\
           & 1 &   6400  &  6.49e-2  &  0.96 \\
           & 1 &  12800  &  3.38e-2  &  0.94 \\
           & 1 &  25600  &  1.80e-2  &  0.91 \\
           & 1 &  51200  &  9.95e-3  &  0.86 \\
\tableline
Adaptive  & 1 &    400  &  5.21e-1  &       \\
          & 2 &    800  &  3.63e-1  &  0.52 \\
          & 3 &   1600  &  2.33e-1  &  0.64 \\
          & 4 &   3200  &  1.26e-1  &  0.89 \\
          & 5 &   6400  &  6.55e-2  &  0.94 \\
          & 6 &  12800  &  3.49e-2  &  0.91 \\
          & 7 &  25600  &  1.90e-2  &  0.88 \\
          & 8 &  51200  &  1.07e-2  &  0.83
\enddata
\tablenotetext{a}{Total levels of the grid}
\tablenotetext{b}{For adaptive meshes, this means the equivalent
  number of zones.}
\end{deluxetable}

Note that the behavior of transverse velocity in \emph{one-dimensional}
Riemann problems is purely due to relativistic effects.  One might have the
following argument which would lead to the wrong conclusion that transverse
velocity cannot change in one-dimensional Riemann problems.  The initial
transverse velocity before the breakup could disappear by transferring to a
new reference frame which moves along the transverse direction in the
original laboratory frame.  In the new frame, no transverse velocity would
develop if the initial transverse velocity in that frame is zero.  After the
breakup of the initial discontinuity, one could transfer back to the original
frame.  The second Lorentz transformation would make the transverse velocity
in the original back to its original value.  Thus the transverse velocity in
the original frame should be fixed during the evolution of the Riemann
problem.  However, relativistic effects invalidate the above argument.
Suppose the original setup is a two-dimensional tube and there is a diaphragm
at the initial discontinuity.  The breakup of the initial discontinuity is
caused by taking the diaphragm away.  This problem can be considered
one-dimensional in the original frame because of the obvious symmetry.
However, in the second reference frame, which moves along the transverse
direction, the problem can no longer be considered one-dimensional.  The
simultaneous disappearance of the diaphragm at different places in the
transverse direction in the original frame does not happen simultaneously in
the second frame.  Therefore, transverse velocity does change in both the
original and the second frame.  This conclusion can also be seen directly
from the governing equations (Eq.~\ref{dudt}) and the definition of the
conserved mass and momentum densities (Eqs.~\ref{d} and \ref{sj}).  These
imply that the SRHD evolution equations preserve $hWv_t$ across shocks and
rarefaction fans, where $v_t$ is a velocity component transverse to the
propagation direction of these waves.  Since, in general, $h$ and $W$ change,
$v_t$ is generally not conserved.

\subsection{Advection Across Jumps in Refinement}

\begin{figure}
%\epsscale{1.0}
%\plotone{advection.eps}
\plotone{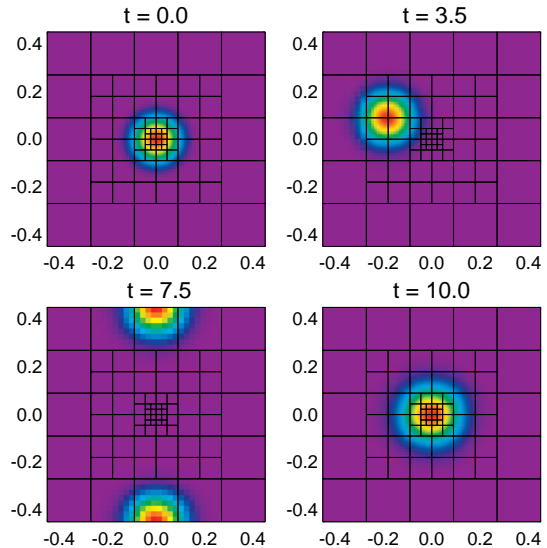}
\caption{Advection across jumps in refinement.  F-WENO with 4 levels
  of refinement fixed in time is used in this calculation.  The block
  structure of the mesh is shown by the black solid lines.  Each block
  contains $8 \times 8$ zones.  Results of the density structure at $t
  = 0.0$ (upper left panel), $t = 3.5$ (upper right panel) $t = 7.5$
  (lower left panel), and $t = 10.0$ (lower right panel) are shown.
  The density pulse advects at a speed of $v = 0.9$ at an angle of
  $\arctan{(4/3)}$ with respect to the $x$-axis and moves across the
  refinement boundaries..  The boundaries of the numerical box are
  periodic.
\label{fig:advection}}
\end{figure}

On AMR grids, there are jumps in resolution at the boundaries between
computational regions at different levels of refinement.  Spurious
effects such as the reflection of waves could arise at these
boundaries.  If AMR performs as desired, the effects should be
negligible because the structure at the refinement boundaries should
be very smooth.  To assess the effects of jumps in refinement in the
worst case scenario, we present an advection test on a fixed staggered
mesh.  We set up a computational grid with a range of refinement
levels, but we turn off adaptive refinement and derefinement so that
the mesh remains spatially variable but constant in time (see
Fig.~\ref{fig:advection}).

% f10

In this test, the computational region consists of a two-dimensional
box in Cartesian coordinates with $-0.45 \le x \le 0.45 $ and $-0.45
\le y \le 0.45 $.  All the boundaries are periodic.  A fixed mesh with
4 levels of refinement is set up on the grid.  On this non-adaptive
staggered mesh (Fig.~\ref{fig:advection}), there is a reference state
of $p = 1.0$, $\rho_\mathrm{ref} = 1.0$, $v_x = 0.72$, $v_y = 0.54$,
with a pulse only in density.  The density profile is given by
$\rho_0(r)$ (Eqs.~\ref{eq:rhocolella} \& \ref{eq:fcolella}), where $r$
is the distance to the center of the box, and the two parameters (the
amplitude and width of the pulse) in the density profile are set to
$\alpha = 10.0$, $L = 0.2$.  The adiabatic index in the equation of
state is $\Gamma = 5/3$ in this test.  The density pulse will advect
on the periodic numerical grid at an angle of $\arctan{(4/3)}$ 
with respect to the $x$-axis and move across the refinement boundaries.
The results are shown in Fig.~\ref{fig:advection}.  As it propagates,
the pulse becomes wider due to numerical viscosity.  However, no
spurious waves are visible, and the pulse is still symmetric.  At $t =
10$, the fluctuation in pressure is only at the level of $10^{-13}$.
Such a small fluctuation, which is probably due to round-off errors in
floating point operations, suggests that spurious effects due to the
jump in refinement is acceptably small in our simulations.

\subsection{Relativistic Jet In Two-Dimensional Cylindrical Geometry}

Many authors have performed two-dimensional simulations to study the
morphology and dynamics of relativistic jets \citep[e.g.,][]{van93,
mar94, dun94, mar97, kom98} mainly in the context of AGNs.  These
results confirmed the formation of the basic features of a jet, i.e.,
beam, cocoon, Mach disk and bow shock, which have been observed in
Newtonian hydrodynamic simulations \citep[e.g.,][]{nor82}.  It was
also found that relativistic jets are more stable and propagate more
efficiently than Newtonian ones.  Three-dimensional relativistic
hydrodynamic simulations of jets have also been performed to study
three-dimensional effects, such as 3D instabilities and precession of
jets \citep{hug02, alo03}.

In this test, we simulate a relativistic jet which is relevant in
astrophysics.  The computational region is a two-dimensional
cylindrical box ($0 \le r \le 15$, $0 \le z \le 45$).  The details of
treating curvilinear coordinates including both cylindrical and
spherical coordinates can be found in appendix~\ref{app}.  For the
sake of comparison, we use the same parameters for this problem as
model C2 of \citet{mar97}.  The initial parameters of the relativistic
jet beam are, $v_b = 0.99$ and $\rho_b = 0.01$.  The classical Mach
number of the jet is set to $M_b = 6$, and the corresponding
relativistic Mach number, $\mathcal{M}_b = W_b M_b /W_s$, where $W_b$
is the Lorentz factor of the jet beam and $W_s$ is the Lorentz factor
of sound speed in the jet, is $\sim 42$.  $\Gamma = 5/3$ is used
for the equation of state.  Initially the computational region is
filled with a uniform medium with $\rho_m = 1$, and $v_m = 0$.  The
pressure of the medium, which is chosen to match the pressure of the
jet, is $p_m = p_b = 0.000170305$.  The initial relativistic jet is
injected through the inlet part ($r \le 1$) of the low-$z$ boundary by
assigning the state of jet material to the boundary.  Outflow boundary
conditions with zero gradients of variables are used at the other part
($r > 1$) of the low-$z$ boundary, high-$z$ boundary, and high-$r$
boundary.  A reflecting boundary is used at the symmetric axis where
$r = 0$.

\begin{figure}
%\epsscale{1.0}
%\plotone{jet_C2.eps}
\plotone{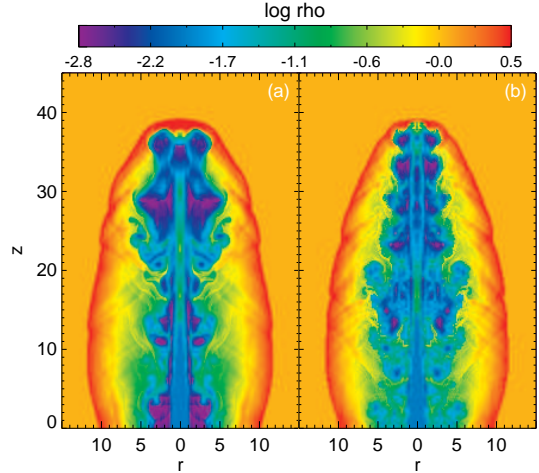}
\caption{Highly supersonic relativistic jet in 2D cylindrical geometry
  at $t = 100$.  F-WENO-A is used in the two calculations with (a) 5
  and (b) 7 levels of refinement.  The equivalent resolutions are (a)
  $384 \times 1152$ and (b) $1536 \times 4608$ zones.  This
  corresponds to a resolution of $\sim 26$ and $\sim 102$ zones per
  jet beam radius, for (a) and (b), respectively.  The CFL number used
  in this test is $0.3$.
\label{fig:jet}}
\end{figure}

Our numerical results at $t = 100$ are shown in Fig.~\ref{fig:jet}.
It is shown that no carbuncle, which is a pathological
phenomenon in some schemes \citep{quirk}, is generated in these
simulations.  We have performed the simulations with two different
resolutions using F-WENO-A.  In both calculations, the lowest level of
the grid consists of $24 \times 72$ zones.  The corresponding spatial
resolution at this level is $\sim 1.6$ zones per jet beam radius.  The
total levels of refinement are 5 and 7, for the two calculations,
respectively.  This corresponds to an equivalent resolution of $\sim
26$ and $\sim 102$ zones per jet beam radius, for the two
calculations, respectively.  The model in this test is highly
supersonic, and relativistic effects from ultra-relativistic motion
dominates those from internal energy \citep{mar97}.  The expected
morphological features of such a relativistic jet are observed.  A bow
shock is formed due to the supersonic motion of the jet.  The medium
is shocked by the bow shock.  The jet beam is slowed down at the Mach
disk and feeds the cocoon.  The shocked jet beam moves sideways and
then even backwards.  The discontinuity between the shocked jet
material and shocked medium material admit Kelvin-Helmholtz
instabilities in the cocoon.  The average speed of the jet head is
$\sim 39$.  This agrees with the essentially one-dimensional analytic
estimate, $42$ \citep{mar97}.  Comparing the results from the two
calculations with different resolutions, we found that the global
structure and the propagation speed of the jet are almost identical
though more mixing due to Kelvin-Helmholtz instabilities is observed
in the high resolution run.

% f11

The presence of the strong shear motion between the jet beam and the
back flows challenges numerical codes.  Even though a smaller CFL
number, 0.3, is used in this test, our code still falls back to more
diffusive schemes occasionally.  We consider this as a signal that the
calculations are still under-resolved for such strong shear flows.  We
note that this is the only test that the fall-back scheme
(\S~\ref{sec:fallback}) is used.

Admittedly, this test problem is not ideal for AMR.  At later times,
most of the region is at the highest refinement level because of the
rich structures inside the bow shock.  However, AMR still saves a lot
of computer time because the external medium requires very low
resolution during most of the time.

\subsection{Relativistic Kelvin-Helmholtz Instability}
\label{sec:kh}

The relativistic Kelvin-Helmholtz instability is of great interest in
many astrophysical problems.  For instance, a key unanswered question
for GRB jets is whether or not the relativistic outflows can remain
sufficiently clean or whether baryons will be mixed into the flow
lowering the maximum asymptotic Lorentz factor below the values
required by observations.  This problem is especially relevant for
collapsars in which the jet must propagate through a star.  The
Kelvin-Helmholtz instability for relativistic jets has been studied
both numerically and analytically \citep[e.g.,][]{ros99,har00,alo02}.
More recently, \citet{per04a,per04b,per05} have performed a series of
studies on the linear growth and long-term nonlinear evolution of
two-dimensional relativistic planar jets.  In this section we present
two numerical tests of the Kelvin-Helmholtz instability.

\begin{figure}
%\epsscale{1.0}
%\plotone{kh.eps}
\plotone{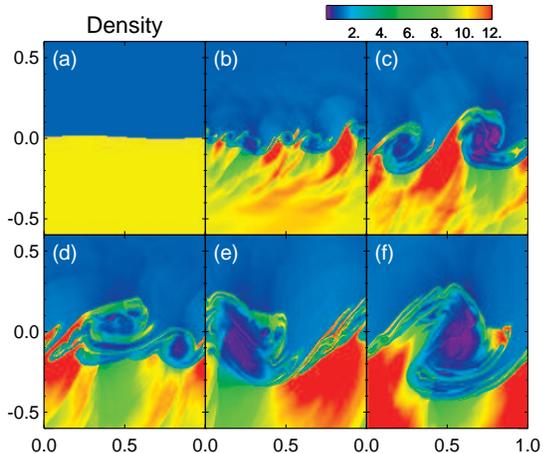}
\caption{Relativistic Kelvin-Helmholtz Instability.  Results at $t =
  0$, $1$, $2$, $3$, $4$, and $5$ are shown in panels (a), (b), (c),
  (d), (e), and (f), respectively.  F-WENO-A with 5 levels of
  refinement is used in this calculation.  The equivalent spatial
  resolution is $1024 \times 10240$ zones.  The whole computational
  region consists of a two-dimensional box with $0 \le x \le 1$ and
  $-5 \le y \le 5$.  In this figure, only part of the domain is shown.
  Initially, the upper half of the fluid moves at a speed of $0.9 c$,
  while the bottom half is at rest.
\label{fig:kh}}
\end{figure}

It is well known that ultra-relativistic flows suppress the
Kelvin-Helmholtz instabilities \citep[see][for a recent analytical
work]{bmr04}.  In the first test in this section, we present a
numerical simulation of the Kelvin-Helmholtz instability with mildly
relativistic speeds to demonstrate the ability of AMR for problems
involving small scale structures.  The computational region consists
of a two-dimensional box with $0 \le x \le 1$ and $-5 \le y \le 5$.
F-WENO-A with 5 levels of refinement is used in this calculation.  The
equivalent spatial resolution is $1024 \times 10240$ zones.  The upper
half of the box is filled with a gas with $\rho = 1$, $p = 1000$, $v_x
= 0.9$ and $v_y = 0$, whereas the bottom half with $\rho = 10$, $p =
1000$, $v_x = 0$ and $v_y = 0$.  $\Gamma = 5/3$ is used for the
equation of state of ideal gas.  Periodic boundary conditions are used
for the right and left boundary.  The initial interface which
separates the two fluids is described by $y = 0.01 \mathrm{sin}(2 \pi
x)$.  This small perturbation triggers the Kelvin-Helmholtz
instability.  Fig.~\ref{fig:kh} shows a series of snapshots of the
results.  The rolling up of the interface forms small vortices.  This
process continues to form larger vortices from smaller vortices and
becomes unstable.

% f12

To allow for a direct comparison with previous results on the
relativistic Kelvin-Helmholtz instability, we repeated model~D10 in
\citet{per05} for a relativistic sheared planar jet using the RAM
code.  In this test, the computational region consists of a
two-dimensional box in Cartesian coordinates with $0 \le x \le 8$ and
$-40 \le y \le 40$.  The jet and ambient medium initially have the
same pressure, $p_0 = 2.0$.  The jet, which moves along the
$x$-direction, has initial density of $\rho_j = 0.1$, and an initial
Lorentz factor of $W_j = 10.0$.  The static ambient medium has initial
density of $\rho_a = 1.0$.  To allow for a continuous transition
between the jet and medium, the profiles of density and velocity are
smoothed as follows,
\begin{equation}
\rho_0(y) = \rho_a - \frac{\rho_a - \rho_j}{\cosh{(y/R_j)^m}}, 
\end{equation}
\begin{equation}
v_x(y) = \frac{v_j}{\cosh{(y/R_j)^m}},
\end{equation}
where the jet radius $R_j$ is set to $1.0$, and the steepness
parameter $m$ is set to $25$.  The adiabatic index in the equation of
state is $\Gamma = 4/3$.  A periodic boundary condition is imposed at
both the low-$x$ and high-$x$ boundary.  To excite the
Kelvin-Helmholtz instability, a small initial perturbation is given to
the transverse velocity, $v_y$:  
\begin{eqnarray}
v_y = \frac{v_1}{N} \left\{ \sum^{N-1}_{n=0}{ \sin{[(n+1)k_nx + \phi_n]}
  \sin^2{[(n+1)\pi y]} \frac{y}{|y|}} \right\}  \nonumber \\  
+ \frac{v_1}{M} \left\{ \sum^{M-1}_{m=0}{ \sin{[(m+1)k_mx + \phi_m]}
  \sin^2{[(m+1)\pi y]}} \right\}, \nonumber \\
\end{eqnarray}
where $v_1 = 5.77 \times 10^{-6}$ (Perucho 2005, private communication)
is the amplitude, $k_{m,n}$ are the wavenumbers of the modes, $\phi_{m,n}$
are the random phases for each mode, $N=4$ is the number of symmetric
modes, and $M=4$ is the number of antisymmetric modes.  The
wavenumbers for these modes are given by $k_j=(j+1) 2 \pi /L $,
where $j=0,1,2,3$, and $L = 8.0$ is the length of the periodic box in
$x$-direction.  

In the test of model D10, 6 levels of mesh refinement are used.  On
the finest level, the resolution is 32 zones and 256 zones per unit
length at $x$ and $y$-direction respectively.  In this test,
resolution is concentrated at the interface between the jet material
and ambient medium.  Instead of using the general refinement criteria
(\S~\ref{sec:amr}), we use the composition of jet material and ambient
medium to determine whether refinement or derefinement should be
performed.  More specifically, the refinement level will be brought to
the highest if the mass fraction of jet material is $0.001 < X_{j} <
0.999$, or otherwise will be marked for derefinement.

\begin{figure}
%\epsscale{1.0}
%\plotone{kh0.eps}
\plotone{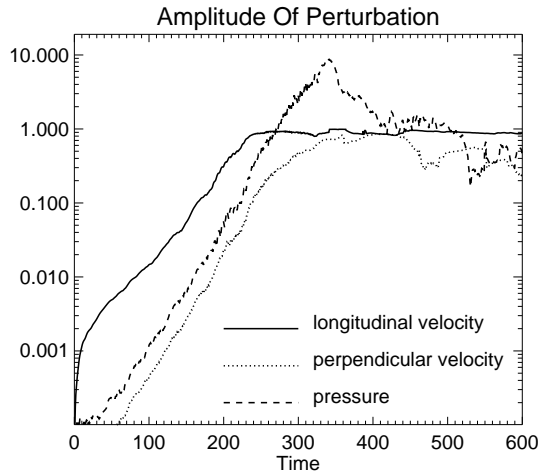}
\caption{Evolution of the relative amplitude of perturbations for
  model~D10 of relativistic Kelvin-Helmholtz instability.  The
  amplitudes of perturbation in the ``jet reference frame'' are defines
  as $0.5 (v_{x,\mathrm{max}}-v_{x,\mathrm{min}}) $, and $ 0.5
  (v_{y,\mathrm{max}}-v_{y,\mathrm{min}}) $, for the longitudinal
  velocity (solid line) and perpendicular velocity (dot line)
  respectively.  The amplitude of the pressure perturbation (dash
  line) is defined as $(p_\mathrm{max}-p_0)/p_0$, where $p_0$ is the
  initial pressure.
\label{fig:kh0}}
\end{figure}

\begin{figure}
%\epsscale{1.0}
%\plotone{kh1.eps}
\plotone{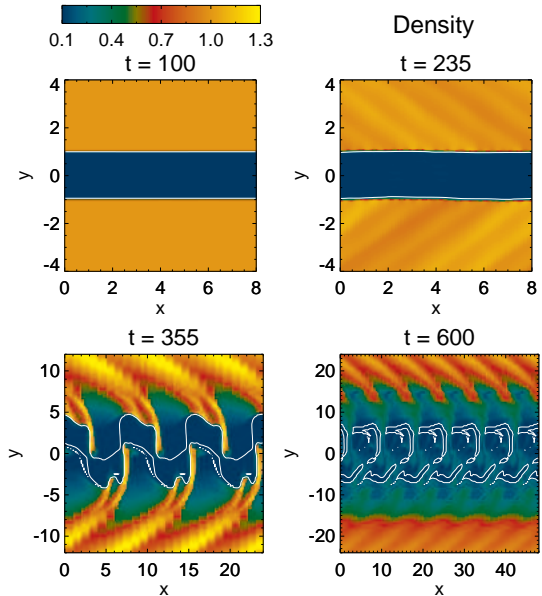}
\caption{Density structure of model~D10 from \citet{per05} of
  the relativistic Kelvin-Helmholtz instability.  F-WENO-A with 6 levels
  of refinement is used in this calculation.  The whole computational
  region consists of a two-dimensional box, $0 \le x \le 8$, $-40 \le
  y \le 40 $, with periodic boundary conditions in the  $x$-direction.
  Results at $t = 100$, 235, 355, and 600 are shown.  The four panels
  show the proper density structure during the middle of the linear regime
  (upper left), at the end of the linear regime (upper right), at the end of the
  saturation regime (lower left), and during the fully non-linear regime (lower
  right), respectively.  The solid white lines denote where the mass
  fraction of the jet material is $X_j = 0.5$.
\label{fig:kh1}}
\end{figure}

Figures~\ref{fig:kh0} \& \ref{fig:kh1} show the results of model D10
using F-WENO-A.  Our results agree with those of \citet{per05}.  The
evolution of the Kelvin-Helmholtz instability consists of three
regimes: linear, saturation and non-linear \citep[see][for more
  details]{per05}.  In the linear regime, the perturbations of the
longitudinal velocity, $v_x$, transverse velocity, $v_y$, and pressure
all grow exponentially (Fig.~\ref{fig:kh0}).  At the end of the linear
regime, $t_\mathrm{lin} \approx 235$, the growth of the longitudinal
velocity perturbation starts to saturate, while both the transverse
velocity and pressure perturbations still grow exponentially.  In the
saturation regime, the continuous growth of the perturbation of
transverse velocity, $v_y$ results in the distortion of the jet beam
(Fig.~\ref{fig:kh1}).  The transition of the saturation to fully
non-linear happens when the pressure perturbation and transverse
velocity perturbation reach their peak at $t_\mathrm{peak} \approx 340$
and $t_\mathrm{sat} \approx 355$ respectively.  In the non-linear regime
the jet material is completely mixed with the ambient medium by turbulent
motions. 

% f13

% f14

\subsection{Emery Step}
\label{sec:emery_amr}

\begin{figure}
%\epsscale{1.0}
%\plotone{emery_amr.eps}
\plotone{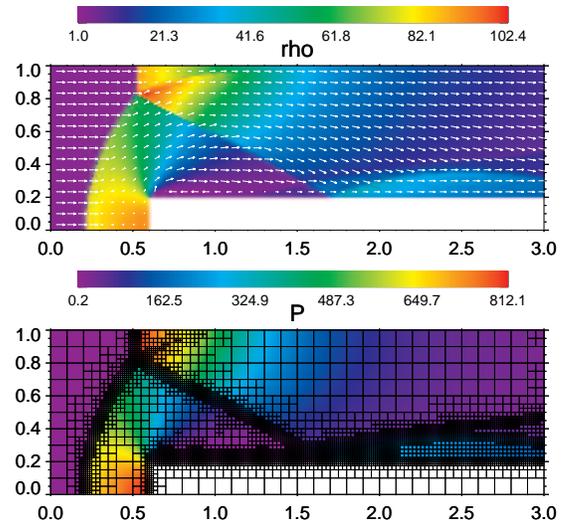}
\caption{ Emery step at $t=4.0$ using F-WENO-A with 5 levels of
  refinement. The equivalent resolution is $\Delta x = \Delta y =
  1/1280$.  Proper mass density and pressure are plotted in the upper
  and bottom panel, respectively.  The velocity field is also plotted
  in the upper panel.  In the bottom panel, the block structure of the
  adaptive mesh is shown.  Each block contains $8 \times 8$ zones. 
\label{fig:emery_amr}}
\end{figure}

The Emery step is a standard test involving the flow in a wind tunnel
encountering a reflecting step boundary.  We have shown the results
with four different schemes (F-WENO, F-PLM, U-PPM, and U-PLM) on
uniform grids in \S~\ref{sec:emery}.  Here, we repeat the test with
F-WENO-A on adaptive meshes.  The setup of this problem is the same as
in \S~\ref{sec:emery}, except that an AMR grid with 5 levels is used
instead of uniform grids.  The equivalent resolution is $3840 \times
1280$ zones on a two-dimensional box of $0 \le x \le 3$ and $0 \le y
\le 1$.  In other words, this corresponds to a zoning of $\Delta x =
\Delta y = 1/1280$.  Again, no special treatment for the step other
than a reflecting boundary is used because we are only interested in
the global structures of shocks.  This problem involves a Mach
reflection of a strong shock by the upper boundary, and then the
reflected shock is reflected again by the step.  AMR should work very
well for this problem because much of the computational volume is
smooth flow requiring only low resolution.  Thus this problem can test
not only the ability of the AMR code to capture strong shocks, but
also its ability to adaptively refine and derefine the mesh as
discontinuities evolve.  It is shown in Fig.~\ref{fig:emery_amr} that
the AMR in our code captures the sharp transitions in the flow where
and when they occur.  Even the weak contact discontinuity originating
from the bottom of the Mach stem is detected and captured by the AMR.
It is interesting to note that no Kelvin-Helmholtz instability
develops along the contact discontinuity though it does for
non-relativistic flow.

% f15

\subsection{Double Mach Reflection Of A Relativistic Shock}

Double Mach reflection of a strong shock has often been used to test
Newtonian hydrodynamics codes since it was first introduced by
\citet{wc84}.  The computational region consists of a two-dimensional
box with $0 \le x \le 4$ and $0 \le y \le 1$.  A reflecting wall is
placed at the $x > 1/6$ part of the low-$y$ boundary.  A strong shock
is moving towards right with its front making a 60 degree angle with
respect to the $x$-axis.  The reflection of the incident shock could
give arise to a double Mach reflection.  Two Mach stems and two
contact discontinuities would appear.  The first contact discontinuity
extends from the first triple shock point to the reflecting wall and
then is pushed by the pressure gradient there and forms a jet parallel
to the wall.

The original setup of the problem cannot be used for special
relativistic hydrodynamics because the specific parameters chosen are
not appropriate for the relativistic case.  To test our relativistic
schemes, we modified the original parameters of the problem.  The
adiabatic index in the equation of state is $\Gamma = 1.4$.  The
unshocked gas has $\rho_0 = 1.4$, $p_0 = 0.0025$, and $v_0 = 0$.  The
classical Mach number of the shock, $M = v_s / c_s$, where $v_s$ is
the speed of the shock front and $c_s$ is the sound speed of the
unshocked gas, is 10.  Using the relativistic shock jump conditions,
we can set up the shocked gas.  The shock front initially intersects
the lower boundary at the edge of the wall, $x = 1/6$.  The $x < 1/6$
part of the low-$y$ boundary and the low-$x$ boundary are set to the
exact post-shock state.  At $y = 1$, the boundary conditions are set
to either post-shock or pre-shock conditions depending upon the exact
motion of the shock.

The post-shock gas moves at a speed of $v_1 \approx 0.4247$, and the
shock speed is $v_s \approx 0.4984$.  We would have made the shock
ultra-relativistic.  However, an ultra-relativistic shock cannot
generate a Mach reflection.  This is qualitatively understandable.
Suppose that the shock front makes an angle of $\theta$ with respect
to the wall and the shock velocity is $v_s$.  The velocity of the
intersection point of the shock and the wall would be
$v_s/\mathrm{sin}(\theta)$.  With ultra-relativistic shock velocity,
the intersection point can move faster than the speed of light.  Note
that this does not violate relativity because the velocity of the
intersection is not a physical velocity.  Even if a Mach stem can be
formed initially, the vertical Mach stem cannot move faster then the
speed of light, and thus cannot follow the motion of the oblique
incident shock.  Therefore no permanent Mach stem is possible when the
shock moves too fast.

\begin{figure}
%\epsscale{1.0}
%\plotone{dmr.eps}
\plotone{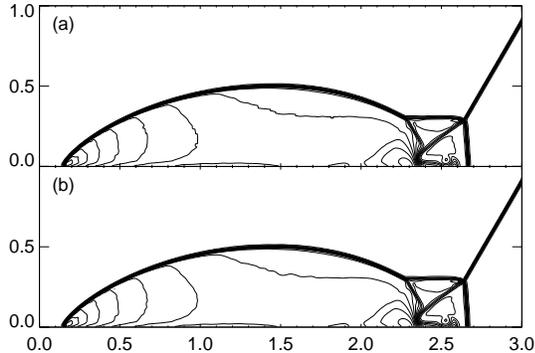}
\caption{Double Mach reflection of a relativistic shock at $t = 4$.
  Results with and without AMR are shown in panels (a) and (b),
  respectively.  Thirty equally spaced contours of proper density are
  plotted.  Only part of the computational domain is shown.  Results
  from (a) F-WENO-A and (b) F-WENO-U are almost identical.
\label{fig:dmr}}
\end{figure}

We have performed this test with both adaptive mesh (F-WENO-A) and
uniform grid (F-WENO-U).  Our results are shown in Fig.~\ref{fig:dmr}.
For the adaptive mesh, 4 levels of refinement are used with the lowest
level containing $64 \times 16$ zones.  The equivalent resolution is $
512 \times 128 $.  For the uniform grid, $ 512 \times 128 $ zones are
used.  The results from F-WENO-A can be compared with those from
F-WENO-U.  The AMR scheme works very well and the highest level of
refinement follows the motion of shocks and discontinuities while most
of the computational region is at lower refinement levels.  It is
shown in Fig.~\ref{fig:dmr} that density contours for F-WENO-A and
F-WENO-U are almost identical.  Our results do not suffer from some
pathological behaviors obtained with some schemes \citep{wc84,quirk}.
There are no kinked Mach stems.  And the region behind the slowly
moving shock near the left edge of the reflecting wall is very smooth.

% f16

\subsection{Three-Dimensional Blast Wave With Spherical Symmetry}
\label{sec:bw3d}

In this test, we study a three dimensional spherical blast wave using
Cartesian coordinates.  Since no analytic solution exists for this
problem, we compare the three dimensional solution to a high
resolution one dimensional simulation run with spherical coordinates.
For the sake of comparison, we perform the same test run by
\citet{db02}.  See their Fig.~8.

The computational domain is a cubic box, $0 \le x \le 1$, $0 \le y \le
1$, $0 \le z \le 1$, with a base resolution of 40 zones in each
direction.  We use up to 4 levels of AMR so the effective resolution
is $320 \times 320 \times 320$ zones.  An initial discontinuity is at
$R = 0.4$, where $R$ is the distance to the center of the Cartesian
coordinates.  The region inside the interface, $R \le 0.4$, contains a
gas with $\rho = 1$, $p = 1000$, whereas the region outside it, $R >
0.4$ contains a gas with $\rho = 1$, $p = 1$.  $\Gamma = 5/3$ is
used for the equation of state.  Both gases are at rest initially.
Similar to the one-dimensional Riemann problem, the decay of the
initial discontinuity will give arise to a spherically symmetric shock
moving outwards, a spherical rarefaction wave moving inwards, and a
``contact discontinuity'' in between.  Note that the system could
eventually evolve into a selfsimilar Blandford-McKee solution for
ultra-relativistic blast waves \citep{bm} if the initial conditions are
modified to make an even stronger explosion.  The boundaries at $x =
0$, $y = 0$ and $z = 0$ are reflecting, whereas all other boundaries
are zero gradient outflows.

\begin{figure}
%\epsscale{1.0}
%\plotone{bw3d.eps}
\plotone{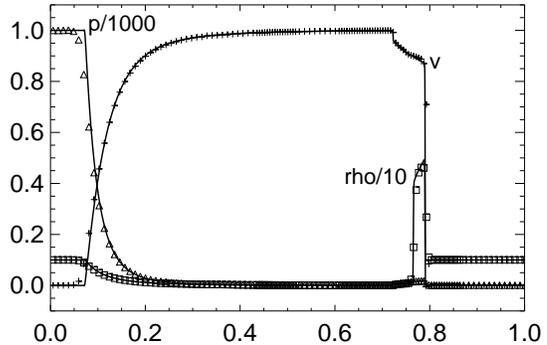}
\caption{Three-dimensional blast wave at $t = 0.4$.  F-WENO-A with 4
  levels of refinement is used for the simulation.  The equivalent
  resolution is $320 \times 320 \times 320$ zones for a cubic box, $0
  \le x \le 1$, $0 \le y \le 1$, $0 \le z \le 1$.  The results along
  the major diagonal of the cubic box are plotted.  Three-dimensional
  numerical results are shown in symbols, whereas the high resolution
  numerical results in one-dimensional spherical coordinates are shown
  in solid lines.  We show proper mass density (square), pressure
  (triangle) and total velocity (plus sign).
\label{fig:bw3d}}
\end{figure}

Fig.~\ref{fig:bw3d} shows the result at $t=0.4$ for a simulation using
a CFL number of 0.3.  The solid line is a one-dimensional simulation
run using spherical coordinates with 4000 uniform zones.  The details
of the treatment of curvilinear coordinates can be found in the
appendix.  The results with three-dimensional Cartesian coordinates
agree with the high resolution results with one-dimensional spherical
coordinates.  The shock front is captured with $\sim 3$ zones.  The
spherical symmetry is kept in the overall structure of the numerical
results using Cartesian coordinates.

% f17

\subsection{Scaling Tests on Parallel Machines}

\begin{figure}
\epsscale{1.0}
%\plotone{efficiency.ps}
\plotone{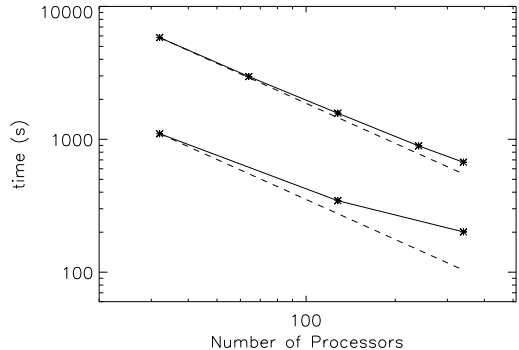}
\caption{ Scaling of two ``constant total work'' parallel simulations
 run on NASA's Columbia supercomputer from 32 to 340 processors.  The
 top pair of lines are for a simulation with a spatially variable but
 temporally fixed mesh.  The bottom pair is for a simulation using
 adaptive refinement.  The dashed lines represent perfect scaling.
\label{fig:efficiency}}
\end{figure}

Many challenging problems in numerical astrophysics require the
computing power made possible by massively parallel supercomputers.
These include state-of-the-art machines available through the national
supercomputing centers and the national laboratories, as well as the
Linux clusters available in an increasing number of institutions.  In
order to take full advantage of available computational power, it is
necessary to develop computer code and utilize packages which run
efficiently on a variety of parallel platforms and scale well with
the number of processors used.  Large simulations, especially in 3D,
require the use of large numbers of parallel processors to run in a
reasonable amount of time.  In order to use parallel resources
efficiently, it is necessary to test how well a code scales with
number of processors used.  Scaling information is important for
planning the size of numerical simulations which are possible to run
efficiently.  It is also impoortant for using an efficient number of
processors for a given job.  We have tested RAM on several major
national supercomputers as well as smaller clusters and find that it
scales very well (Figure~\ref{fig:efficiency}).

% f18

RAM has been extensively tested on NASA's Columbia supercomputer on up
to 504 processors. A scaling analysis for a stellar collapse
simulation run with a spatially variable temporally fixed grid shows
how RAM scales from 32 to 340 processors (Figure~\ref{fig:efficiency},
upper lines).  For this constant total work job, RAM runs at 93\%
maximum efficiency when going from 32 to 128 processors and 88\%
maximum efficiency when going from 128 to 340 processors.  We define
efficiency when going from $m$ to $n$ processors as $e \equiv
t_{m}/t_{n} * (m/n)$, where $t_{x}$ is the run time with $x$
processors.  In a different simulation of a relativistic shock using a
spatially and temporally adaptive mesh (Figure~\ref{fig:efficiency},
lower lines), processing time scales with 80\% of maximum efficiency
when going from 32 to 128 processors, and 65\% efficiency going from
128 to 340 processors on Columbia.  This simulation required more
communication among processors due to the continual refinement and
derefinement as flow features moved through the computational domain.

Since the RAM code is based on FLASH 2.3, we expect the scaling and
efficiency to be similar.  More details on the scaling and efficiency
of the FLASH code can be found in \citet{flash}.

\section{Summary and Discussion}
\label{sec:summary}

We have presented a new special relativistic hydrodynamics code with
adaptive mesh refinement.  The code is modular and includes four
combinations of reconstruction and hydrodynamics solvers we have
termed F-WENO, F-PLM, U-PPM and U-PLM.  In this paper, we focus
attention on F-WENO, a characteristic-wise, finite difference WENO
scheme utilizing the full characteristic decomposition of the SRHD
equations.  The scheme is fifth-order accurate in smooth regions.
This is the first time that this high-order scheme has been
implemented for relativistic hydrodynamics.  We demonstrate that,
while somewhat more complex, this method is highly accurate and
stable.  It has the added advantage of not containing tunable
parameters which substantially modify the performance of the
algorithm.

The U-PPM scheme utilizes an approximate Riemann solver (e.g., modified
Marquina flux formula) and is equivalent to the GENESIS code \citep{alo99}.
U-PLM and F-PLM, use linear reconstruction and are more diffusive, though
somewhat stabler, versions of U-PPM and F-WENO.  For some problems these more
diffusive schemes perform better when suppression of spurious oscillations
near discontinuities is desirable.  The higher-order methods sometimes
produce numerical oscillations near strong discontinuities leading to values
of conserved variables which are inconsistent with primitive fluid variables.
This inconsistency can lead to code crashes.  We therefore have implemented a
failsafe scheme, which saves the previous solution at all times.  If a
numerical step results in an unacceptable solution, we return to the previous
step and repeat the hydrodynamics calculation with increasingly diffusive
schemes which, due to their diffusive nature, are more likely to produce
consistent variables.  This procedure greatly increases code robustness.

Numerical relativistic hydrodynamics is more difficult than the Newtonian
case for several fundamental reasons.  First, due to relativistic effects,
structures in relativistic flows tend to be thinner requiring higher spatial
resolution.  Second, unlike shocks in Newtonian hydrodynamics, the density
jump across relativistic shocks can be arbitrarily large and is limited only
by the Lorentz factor (see Eq.~\ref{sigma}).  Third, the non-linear coupling
between the velocity components in relativistic flows remains a very
challenging numerical problem.  The Lorentz factor depends on all velocity
components and couples transverse velocity into the dynamics in the normal
direction.  The difficulty is in resolving contact discontinuities which move
with respect to the numerical grid.  Some smearing of contact discontinuities
is inevitable due to numerical diffusion error.  When transverse velocity
components are present they experience jumps across contact discontinuities
with corresponding jumps in the Lorentz factor, unlike the case with no
transverse velocity in which the Lorentz factor is continuous across contact
discontinuities.  Because the conserved variables depend on different powers
of the Lorentz factor, the numerically smeared states generate spurious waves
from the discontinuity which corrupt the solution. An additional, though less
serious, difficulty is that pressure jumps across the shock and inside the
rarefaction fan have a steeper dependence on normal velocity when transverse
velocities are present.

These problems can be solved by increasing spatial resolution.  A major
advantage of RAM is its high accuracy coupled with AMR for extremely high
effective resolution.  We have found that RAM is able to achieve sufficient
resolution to correctly solve challenging numerical problems.  Since AMR
concentrates the spatial resolution where it is needed, sufficient accuracy is
achieved efficiently, enabling challenging multi-dimensional simulations to be
undertaken.

The high accuracy achieved with RAM is demonstrated to be of critical
importance for solving relativistic flows with strong shear, even in the more
challenging case when transverse velocity is present in high pressure states
into which rarefaction fans are propagating.  Under-resolved simulations with
many combinations of reconstruction and hydrodynamics solvers produce
incorrect post-shock values and positions as seen in Fig.~\ref{fig:rievy2}.
We believe this problem is generic to most or all relativistic hydrodynamics
codes currently in use \citep[see also][]{mig05}.  In this work, we succeed
in obtaining agreement with the analytic solution for a Riemann problem with
transverse velocity of $v = 0.9$ by using 8 levels of AMR.  We are confident
that we can use RAM to adequately resolve flows with strong shear which have
not previously been adequately resolved.

It is worth noting that U-PPM and U-PLM schemes do not require the complete
characteristic information of the Jacobian matrix of the SRHD system.
However, this information is needed for the characteristic-wise schemes
(F-WENO and F-PLM) making them more difficult to implement.  Our numerical
experiments have shown that the extra effort is justified.  We have also
implemented other F-X schemes for comparison including the fifth-order
monoticity-preserving scheme (MP5) of \citep{mp5} and third-order WENO
(WENO3) \citep{jia96}.  In our numerical experiments we have found that MP5
generates excessive spurious numerical oscillations, while WENO3 is
exceedingly diffusive.  The fifth-order WENO scheme implemented in this work
(see \S~\ref{sec:f-x}) is free from these liabilities.

In order to simulate systems of astrophysical interest we have included
physics modules relevant for the study of GRBs, SN and black hole accretion.
RAM includes physical equations of state, neutrino cooling and nuclear
physics.  It is intended initially to address open theoretical questions in
the study of gamma-ray bursts and supernova explosions.

\acknowledgments

W.Z. has been supported by NASA through Chandra Postdoctoral
Fellowship PF4-50036 awarded by the Chandra X-Ray Observatory Center,
and the DOE Program for Scientific Discovery through Advanced
Computing (SciDAC; DE-FC02-01ER41176).  A.M. acknowledges support at
the Institute for Advanced Study from the Keck Foundation.  We
acknowledge NASA Theory grant NAG5-12036 for the October, 2003 GRB
meeting at UC Santa Cruz during which this work was initiated and NASA
Grant SWIF03-0000-0060 for travel funds supporting this collaboration.
We thank the Institute for Nuclear Theory at University of Washington
for hospitality during part of the time over which this work was
completed.  We acknowledge the use of the UCSC UpsAnd cluster
supported by an NSF MRI grant AST-0079757, the Scheides cluster at the
Institute for Advanced Study, NASA's Columbia computer and resources
of the National Energy Research Scientific Computing Center.  The
software used in this work was in part developed by the DOE-supported
ASCI/Alliance Center for Astrophysical Thermonuclear Flashes at the
University of Chicago.  Specifically, we use the PARAMESH AMR and I/O
tools from FLASH version 2.3.  The hydrodynamics code is original to
this work.  We would like to thank M. Zingale for many helpful
conversations, especially on the FLASH code, J. Stone for many
detailed and useful comments on the manuscript and the anonymous
referee for his/her thorough review.

\begin{appendix}

\section{Special Relativistic Hydrodynamics In Curvilinear Coordinates}
\label{app}

When a method which is valid in Cartesian coordinates is extended to
curvilinear coordinates care must be taken to avoid the introduction
of errors related to the mesh geometry.  We treat the extension to
non-Cartesian coordinates carefully to avoid errors especially near
coordinate singularities.  Here we describe the extension of our code
to cylindrical and spherical coordinate geometries.

In cylindrical coordinates ($r$, $\theta$, $z$), the governing SRHD
equation can be written as,
\begin{eqnarray}
\frac{\partial{D}}{\partial{t}} + \frac{1}{r}
\frac{\partial{}}{\partial{r}} (r D v_r) + \frac{1}{r}
\frac{\partial{}}{\partial{\theta}} (D v_{\theta}) +
\frac{\partial{}}{\partial{z}} (D v_z) & = & 0 \label{eqn:c0} \\
\frac{\partial{S_r}}{\partial{t}} + \frac{1}{r}
\frac{\partial{}}{\partial{r}}\{r(S_r v_r + p)\} +
\frac{1}{r}\frac{\partial{}}{\partial{\theta}}(S_r v_\theta) +
\frac{\partial{}}{\partial{z}} (S_r v_z) & = & 
   \lefteqn{\frac{p}{r} + \frac{\rho h W^2 v^2_\theta}{r}} \label{eqn:c1} \\
\frac{\partial{S_\theta}}{\partial{t}} + \frac{1}{r}
\frac{\partial{}}{\partial{r}}(r S_\theta v_r) + \frac{1}{r}
\frac{\partial{}}{\partial{\theta}}(S_\theta v_\theta + p) +
\frac{\partial{}}{\partial{z}}(S_\theta v_z) & = & 
   \lefteqn{- \frac{\rho h W^2 v_r v_\theta}{r}} \label{eqn:c2} \\
\frac{\partial{S_z}}{\partial{t}} + \frac{1}{r}
\frac{\partial{}}{\partial{r}} (r S_z v_r) + \frac{1}{r}
\frac{\partial{}}{\partial{\theta}} (S_z v_\theta) +
\frac{\partial{}}{\partial{z}} (S_z v_z + p) & = & 0 \label{eqn:c3} \\
\frac{\partial{\tau}}{\partial{t}} + \frac{1}{r}
\frac{\partial{}}{\partial{r}} \{r(S_r - D v_r)\} + \frac{1}{r}
\frac{\partial{}}{\partial{\theta}} (S_\theta - D v_\theta) +
\frac{\partial{}}{\partial{z}} (S_z - D v_z) & = & 0 \label{eqn:c4}
\end{eqnarray} 
where the subscripts, $r$, $\theta$ and $z$ stand for radial,
azimuthal and axial directions in cylindrical coordinates.  All fluid
variables have the same meaning as in \S~\ref{sec:eqns}.

The above equations for cylindrical coordinates can be easily
discretized into, 
\begin{equation}
\frac{d\mathbf{U}_{i,j,k}}{dt}  =  -
    \frac{r_{i+1/2}\mathbf{F}^r_{i+1/2,j,k} -
    r_{i-1/2}\mathbf{F}^r_{i-1/2,j,k}}{r_i \Delta r} -
    \frac{\mathbf{F}^\theta_{i,j+1/2,k} -
    \mathbf{F}^\theta_{i,j-1/2,k}}{r_i \Delta \theta} 
     - \frac{\mathbf{F}^z_{i,j,k+1/2} 
      - \mathbf{F}^z_{i,j,k-1/2}}{\Delta z} + \mathbf{S}_{i,j,k},
\end{equation} 
where the subscripts i, j and k denote the $r$-, $\theta$- and
$z$-discretization, respectively.  The subscripts $i\pm1/2$,
$j\pm1/2$, and $k\pm1/2$, refer to cell interfaces.
$\mathbf{U}_{i,j,k}$ is the mean value of the conserved variable at
the cell ($i$,$j$,$k$).  $\mathbf{F}$ are the numerical fluxes at the
cell interfaces. The source term, $\mathbf{S}_{i,j,k}$, can be
calculated according to the right-hand sides of
Equations~\ref{eqn:c0}, \ref{eqn:c1}, \ref{eqn:c2}, \ref{eqn:c3} and
\ref{eqn:c4}.
  
Similarly, the governing equation in spherical coordinates ($r$,
$\theta$, $\phi$) can be written as,
\begin{equation}
\frac{\partial{D}}{\partial{t}} + \frac{1}{r^2}
\frac{\partial{}}{\partial{r}} (r^2 D v_r) + \frac{1}{r \sin{\theta}}
\frac{\partial{}}{\partial{\theta}} (\sin{\theta} D v_{\theta}) +
\frac{1}{r \sin{\theta}} \frac{\partial{}}{\partial{\phi}} (D v_\phi)
= 0
\end{equation}
\begin{equation}
\frac{\partial{S_r}}{\partial{t}} + \frac{1}{r^2}
\frac{\partial{}}{\partial{r}}\{r^2(S_r v_r + p)\} + \frac{1}{r
\sin{\theta}}\frac{\partial{}}{\partial{\theta}}(\sin{\theta}S_r
v_\theta) + \frac{1}{r \sin{\theta}} \frac{\partial{}}{\partial{\phi}}
(S_r v_\phi)  = \frac{2p}{r} +
\frac{\rho h W^2 (v^2_\theta + v^2_\phi)}{r}
\end{equation}
\begin{equation}
\frac{\partial{S_\theta}}{\partial{t}} + \frac{1}{r^2}
\frac{\partial{}}{\partial{r}}(r^2 S_\theta v_r) + \frac{1}{r
\sin{\theta}}
\frac{\partial{}}{\partial{\theta}}\{\sin{\theta}(S_\theta v_\theta +
p)\} + \frac{1}{r \sin{\theta}}
\frac{\partial{}}{\partial{\phi}}(S_\theta v_\phi)  =
\frac{\cot{\theta}p}{r} + \frac{\rho h W^2 (v^2_\phi \cot{\theta} -
v_r v_\theta)}{r}
\end{equation}
\begin{equation}
\frac{\partial{S_{\phi}}}{\partial{t}} + \frac{1}{r^2}
\frac{\partial{}}{\partial{r}} (r^2 S_\phi v_r) + \frac{1}{r
\sin{\theta}} \frac{\partial{}}{\partial{\theta}} (\sin{\theta}S_\phi
v_\theta) + \frac{1}{r \sin{\theta}} \frac{\partial{}}{\partial{\phi}}
(S_\phi v_\phi + p) 
= - \frac{\rho h W^2 v_\phi (v_r + v_\theta \cot{\theta})}{r}
\end{equation}
\begin{equation}
\frac{\partial{\tau}}{\partial{t}} + \frac{1}{r^2}
\frac{\partial{}}{\partial{r}} \{r^2(S_r - D v_r)\} + \frac{1}{r
\sin{\theta}} \frac{\partial{}}{\partial{\theta}} \{ \sin{\theta}
(S_\theta - D v_\theta)\}  + \frac{1}{r \sin{\theta}}
\frac{\partial{}}{\partial{\phi}} (S_\phi - D v_\phi) = 0,
\end{equation}
where the subscripts, $r$, $\theta$ and $\phi$ stand for radial, polar
and azimuthal directions in spherical coordinates.  All fluid
variables have the same meaning as in \S~\ref{sec:eqns}.

The discretized equations for spherical coordinates read
\begin{eqnarray}
\frac{d\mathbf{U}_{i,j,k}}{dt} = -
    \frac{r^2_{i+1/2}\mathbf{F}^r_{i+1/2,j,k} -
    r^2_{i-1/2}\mathbf{F}^r_{i-1/2,j,k}}{r^2_i \Delta r} 
    - \frac{\sin{\theta_{j+1/2}}\mathbf{F}^\theta_{i,j+1/2,k} -
    \sin{\theta_{j-1/2}}\mathbf{F}^\theta_{i,j-1/2,k}}{r_i \sin{\theta_j}
    \Delta \theta} \nonumber \\
    - \frac{\mathbf{F}^\phi_{i,j,k+1/2} - \mathbf{F}^\phi_{i,j,k-1/2}}
    {r_i \sin{\theta_j} \Delta \phi} + \mathbf{S}_{i,j,k},
\end{eqnarray} 
where the subscripts i, j and k denote the $r$-, $\theta$- and
$\phi$-discretization, respectively.  The subscripts $i\pm1/2$,
$j\pm1/2$, and $k\pm1/2$, refer to cell interfaces.

\end{appendix}

{}

\end{document}